\documentclass[preprint]{vgtc}                

\graphicspath{{figures/}{pictures/}{images/}{./}} 

\usepackage{times}                     

\usepackage{tabu}                      
\usepackage{booktabs}                  
\usepackage{lipsum}                    
\usepackage{mwe}                       
\usepackage{subcaption}                

\usepackage{mathptmx}                  

\onlineid{0}

\vgtccategory{Research}

\vgtcinsertpkg

\preprinttext{\parbox{0.9\textwidth}{\centering\tiny
Accepted for publication at IEEE ISMAR 2026.\\
\copyright~2026 IEEE. Personal use of this material is permitted. Permission from IEEE must be obtained for all other uses, in any current or future media, including reprinting/republishing this material for advertising or promotional purposes, creating new collective works, for resale or redistribution to servers or lists, or reuse of any copyrighted component of this work in other works.}}

\title{Switched Reading: Toward Seamless Visual-Auditory Switching \\When Reading Text in Augmented/Mixed Reality}

\author{Kazuyuki Fujita\thanks{e-mail: k-fujita@riec.tohoku.ac.jp} %
\and Yuto Matsui %
\and Ikuru Sato %
\and Guanghan Zhao\thanks{e-mail: zhao.guanghan.b6@tohoku.ac.jp} %
\and Yoshifumi Kitamura\thanks{e-mail: kitamura@riec.tohoku.ac.jp}}

\affiliation{\scriptsize Research Institute of Electrical Communication, Tohoku University}

\teaser{
  \centering
  \includegraphics[width=\linewidth]{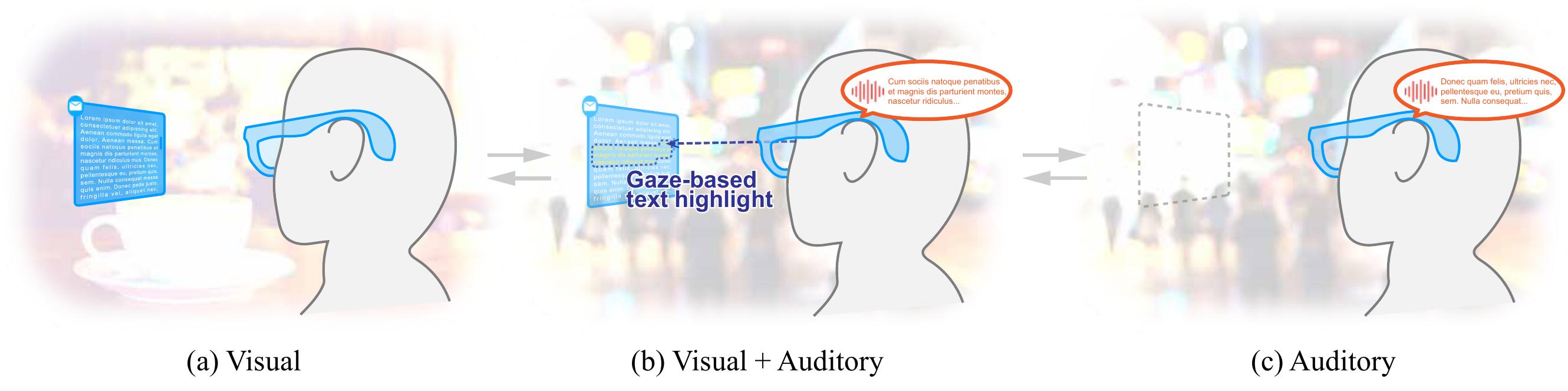}
  \caption{\textit{Switched Reading} allows the user to continue reading text while switching between visual and auditory modalities according to the user's context changes (e.g., from sitting in a cafe (a) to walking around town (b-c)). The interface provides a visual display (a) and voice reading (c) of text, as well as an intermediate state (b) that highlights the currently read content, based on gaze, in the transition between the two modalities.}
  \label{fig:teaser}
}

\abstract{
    Augmented/mixed reality (AR/MR) wearable glasses now permit information interaction anywhere, but visual displays can be inappropriate when real-world awareness is essential. We propose \textit{Switched Reading}, a novel interaction framework for reading text in AR/MR that supports switching between visual and auditory modalities as needed. Specifically, we explore two key interaction techniques within this framework: (1) gaze-based voice playback and (2) a correspondence-aware transition effect. We implemented them on an MR headset through a parameter-tuning user test. Next, we conducted a user study (N=16) to investigate the impact of the two techniques on reading performance and overall user experience with simulated modality switching in virtual reality. The results show that the condition combining both techniques was the most preferred among four conditions. Moreover, we found that the gaze-based voice playback reduced gaze offsets when switching modalities and improved reading speed over the baseline condition using scroll position. Finally, we implemented a Switched Reading application for reading while walking and collected user feedback, yielding further design implications for practical use.
} 


\begin{document}


\firstsection{Introduction}

\maketitle

Recently, the proliferation of head-worn displays (HWDs) has enabled users to interact with computers regardless of location or context. Commercially available augmented/mixed reality (AR/MR) HWDs allow users to display virtual content at arbitrary positions in mid-air without physical constraints. This capability has been explored for various scenarios where traditional displays are inconvenient to use, such as during walking or commuting \cite{reading-onSmartGlass, reading-spacing, comparing-coordinate, dynamicTextManagement, lu2020glanceable, glassMessaging, glassMail, ku2019peritext}, cycling \cite{hazardSnap, zhao2024rear, dancu2015gesture, chatterjee2020smarthelm, matviienko2022bikear}, and exercising (e.g., dancing \cite{here&now} or yoga \cite{flowAR}).

However, interacting with virtual content in AR/MR environments diverts attention away from the real world. Specifically, virtual content displayed in mid-air may occlude important information in the real-world environment or reduce the user's situational awareness, leading to communication breakdowns \cite{Koelle2015Dont, Rzayev2020Effects} or physical risks (e.g., collisions with people or objects while walking) \cite{Huang2024Reading}. Prior research has attempted to address these issues through improvements in content layout \cite{reading-onSmartGlass, reading-spacing, chua2016positioning, orlosky2014managing, dynamicTextManagement}, presentation style \cite{reading-onSmartGlass, ku2019peritext, dancu2015gesture}, text and background colors \cite{textReadability, gabbard2007active}, coordinate systems \cite{comparing-coordinate}, font types \cite{matsuura2019readability}, and interaction timing \cite{lu2020glanceable}. Nevertheless, all of these approaches require users to visually focus on the content, which does not fundamentally resolve the above issues. One task where this is particularly problematic is reading for comprehension, where users must maintain visual attention on the content for extended periods, exacerbating the lack of situational awareness in the real world.

To address this issue, we focus on auditory presentation as an alternative modality and explore a novel way of comprehending text by seamlessly switching between visual and auditory modalities, which we call \textit{Switched Reading}. While the auditory modality (i.e., listening to text) is slower than the visual modality (i.e., reading text) in terms of cognitive processing speed, it is considered more effective in reducing cognitive load in scenarios requiring attention to the external environment, such as walking or driving \cite{reading-onthego,road2productivity}. Accordingly, we believe that Switched Reading has the potential to support users in continuing to read in AR/MR while maintaining their attention, even under changing external contexts. Although a few prior studies have explored similar ideas that involve modality switching while reading (e.g., for smartphones \cite{continuous-reading}), there has been little investigation into detailed user interface design to minimize potential context loss during modality switching. However, many recent HWDs have been equipped with eye trackers, which could be used to achieve smarter switched reading experiences.

Therefore, in this study, we explore the detailed interface design of a system that manages the visual-auditory transition to provide \textit{seamless} Switched Reading in AR/MR (Figure \ref{fig:teaser}). Here, ``seamless'' refers to the ability to maintain the context of the text before and after modality switching while minimizing wasteful repetition. To achieve this, we introduce two key interaction techniques: (1) gaze-based voice playback, where voice reading begins near the reading position estimated from the user's gaze (Figure \ref{fig:teaser}a$\to$b), and (2) a correspondence-aware transition effect, where the interface highlights the currently read content during the visual-auditory modality switching (Figure \ref{fig:teaser}a$\to$b$\to$c and c$\to$b$\to$a). These techniques are expected to reduce the cognitive load required for users to locate their reading position, thus providing benefits in scenarios where visual information presentation is intermittently unavailable or inappropriate (e.g., walking, cycling, exercising, or closing one's eyes).

We developed a prototype of the proposed interface using Meta Quest Pro, an HWD equipped with an eye tracker. Detailed parameters of its behavior were determined through a preliminary user study (N=8). Using this prototype, we conducted a larger user study (N=16) to investigate the impact of our key techniques---(1) gaze-based playback and (2) the correspondence-aware transition effect---on reading performance and user experience during Switched Reading. The results show that gaze-based switching reduced the number of gaze shifts required to locate the reading position and improved reading speed; furthermore, an interface combining both techniques was found to be the type most preferred by users. 
Finally, we implemented a Switched Reading application for reading while walking and collected user feedback.
Based on these findings, we derived design guidelines for interfaces that support a better Switched Reading experience.

The main contributions of this paper are as follows:
\begin{itemize}
    \item Formulation of Switched Reading as an interaction design space for maintaining textual context across visual and auditory presentation in AR/MR, together with a discussion of its possible interaction scenarios,
    \item Design exploration of a Switched Reading interface featuring gaze-based playback and a transition effect that enables seamless modality transitions, along with its prototype running on an MR headset,
    \item A controlled user study (N=16) in VR revealing that our interface's gaze-based voice playback feature contributed to reduced gaze shifts during modality switching and increased reading speed, and
    \item An application study (N=12) of reading while walking, yielding design implications for practical use.
\end{itemize}

\section{Related Work}\label{chap:related_work}

\subsection{Interacting with Floating Content in AR/MR}
In AR/MR, users can access virtual content in mid-air while maintaining awareness of their real-world surroundings~\cite{klose2019text, reading-onSmartGlass, ginters2019augmented}. Thus, an increasing number of research works have examined arranging virtual content in mobile or multitasking contexts~\cite{ghosh2020eyeditor, lucero2014notifeye, orlosky2014managing, reading-onSmartGlass, lu2020glanceable}. For instance, Lu et al.~\cite{lu2020glanceable} proposed \textit{Glanceable AR}, an interaction paradigm in which virtual content is arranged at the edge of the user's field of view to be accessed by glancing at it. Their follow-up studies ~\cite{lu2021evaluating,Lu2023InTheWild} have demonstrated its efficacy in daily life scenarios. However, this approach assumes content that can be grasped instantly (e.g., icons or widgets), and thus it is not well-suited to reading long-form text, the focus of this study.

Furthermore, there is a growing body of research on mobile reading tasks during multitasking (primarily walking)~\cite{reading-onSmartGlass, reading-spacing, comparing-coordinate, dynamicTextManagement}, similar to our study. For example, Rzayev et al.~\cite{reading-onSmartGlass} investigated display positions and methods for text, revealing that placing text in the upper-right field of view increased cognitive load and reduced comprehension. Zhou et al.~\cite{reading-spacing} showed that increasing line spacing improved both reading and walking speed, facilitating smoother task switching between reading and navigation.
However, in such multitasking environments, users must divide their visual attention between the virtual content and the real world, potentially reducing task performance~\cite{oulasvirta2005interaction}. Therefore, our study focuses on the auditory modality as an alternative for information acquisition, aiming to improve the efficiency of content consumption in AR/MR environments by adaptively switching between visual and auditory modalities depending on the user's context.

\subsection{Visual/Auditory Reading}

While both visual and auditory modalities can be used to process text, they differ in their processing characteristics and cognitive efficiency. A representative meta-analysis by Virginia et al.~\cite{listening-or-reading} compared comprehension between visual and auditory modalities across 46 studies. They found that visual reading is more effective when deep understanding is required or when users can control the pace of reading. Moreover, visual modalities provide advantages such as better overview, easy backtracking, and integration with structural elements like figures and tables. These attributes explain why most text-based information acquisition in daily life is visually oriented.

However, auditory presentation is more appropriate in some situations. Vadas et al.~\cite{reading-onthego} compared smartphone-based visual displays and text-to-speech displays for static and walking conditions. They found auditory presentation to be effective for reading while walking, since it reduces cognitive load related to monitoring the external environment. Similar findings have been reported for driving scenarios~\cite{road2productivity}. This aligns with the common practice of listening to the radio while driving, whereas the visual use of smartphones in such contexts is discouraged for safety.

In addition, the simultaneous use of visual and auditory modalities has been explored. Studies have reported that dual-modality presentation can enhance text comprehension compared to using a single modality~\cite{schiavo2021attention, liu2019modality}. However, some users find the auditory reading pace limiting or distracting when accompanied by visual content~\cite{schiavo2021attention}. Consequently, each modality has its own advantages and limitations, suggesting that adaptive switching depending on the context would be ideal.

\subsection{Visual-Auditory Modality Switching}
\label{subsec:related3}
With the widespread adoption of audiobooks and podcasts, auditory presentation has become one of the major modalities for acquiring textual information. While much of the research on auditory text presentation has focused on accessibility~\cite{auralBrowsing, auralNavigation, rohani2016semi}, relatively few studies have investigated the use of both visual and auditory modalities in content acquisition.

One notable example is the study of Hsieh et al.~\cite{visual-audioHaptic}, which proposed an interaction technique that allows users to switch modalities between visual (displaying text on a tablet) and auditory (reading by voice) presentation depending on the situation during a city navigation task. However, their system does not support seamless switching of modalities in the course of content acquisition---an important distinction from our study.

Although not designed for HWDs, Yu~\cite{continuous-reading} proposed a smartphone interface that automatically switches between visual and auditory modalities depending on whether the user is walking or stationary. This system was shown to be more effective than using a single modality. However, it initiates voice playback from the beginning of the currently displayed screen, regardless of the user’s reading position in the text, which may result in a gap between what the user last saw and what they hear. Therefore, it is not possible to guarantee context continuity during switching. In contrast, our study explores a user interface that explicitly supports the maintenance of contextual continuity during modality switching.

\begin{figure*}[t]
    \centering
    \includegraphics[width=1.0\textwidth]{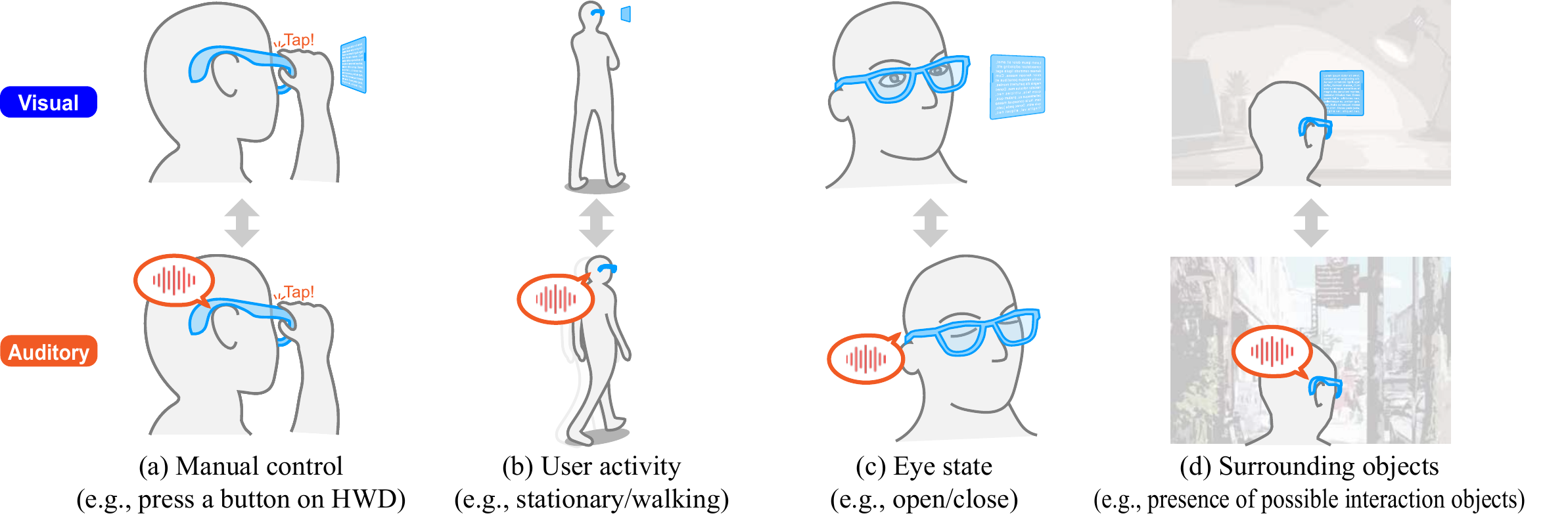}
    \caption{Interaction scenarios of Switched Reading.}
    \label{fig:interaction_scenario}
\end{figure*}

\section{Switched Reading}\label{sec:proposal}
As a practical step toward achieving such an interface, we propose \textit{Switched Reading}, a novel interaction framework for reading text in AR/MR that supports switching between visual and auditory modalities as needed. To support users in maintaining their reading context even during modality switching, we introduce and explore two key interaction techniques within this framework: (1) gaze-based voice playback and (2) a correspondence-aware transition effect. The remainder of this section describes the detailed design of these techniques and the implementation of our prototype, as well as possible interaction scenarios using our framework.

\subsection{Key Interaction Techniques}
Toward seamless switching between visual and auditory modalities while maintaining textual context, we introduce two key techniques as follows.

\noindent\textbf{(1) Gaze-based voice playback}. When switching from visual to auditory presentation, there is inevitably a gap between the user's current visual reading position in the text and the audio playback's starting point, which should be minimized. To deal with this, our interface uses eye gaze: The interface estimates the user's real-time reading position based on gaze direction and starts the audio playback accordingly.

\noindent\textbf{(2) Correspondence-aware transition effect}. To support the user's awareness of the correspondence between the modalities' contents during visual-auditory switching, our interface employs a visual-auditory transition effect. Using this effect, the system visually highlights the audibly read part of the text for a certain amount of time as the modality is switched. This helps users find where to start reading in the other modality, thus maintaining contextual continuity between modalities.

\begin{figure*}[t]
    \centering
    \includegraphics[width=1.0\textwidth]{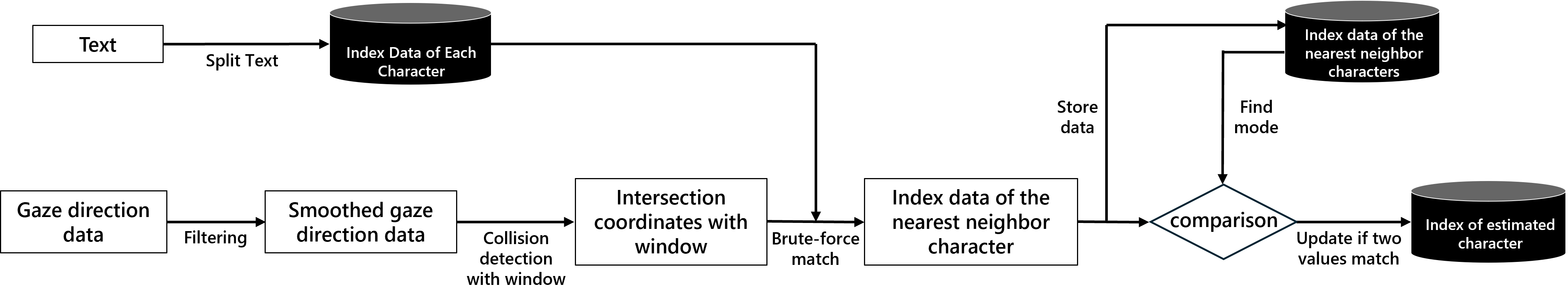}
    \caption{Flow of reading position estimation.}
    \label{fig:RPE}
\end{figure*}

\begin{figure*}[t]
    \centering
    \includegraphics[width=0.9\textwidth]{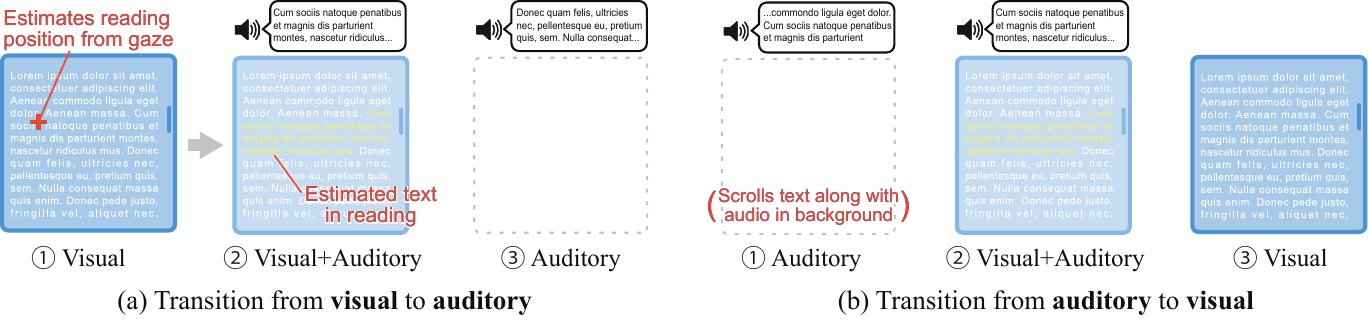}
    \caption{System overview of proposed prototype.}
    \label{fig:System}
\end{figure*}

\begin{figure*}[t]
    \centering
    \includegraphics[width=0.65\textwidth]{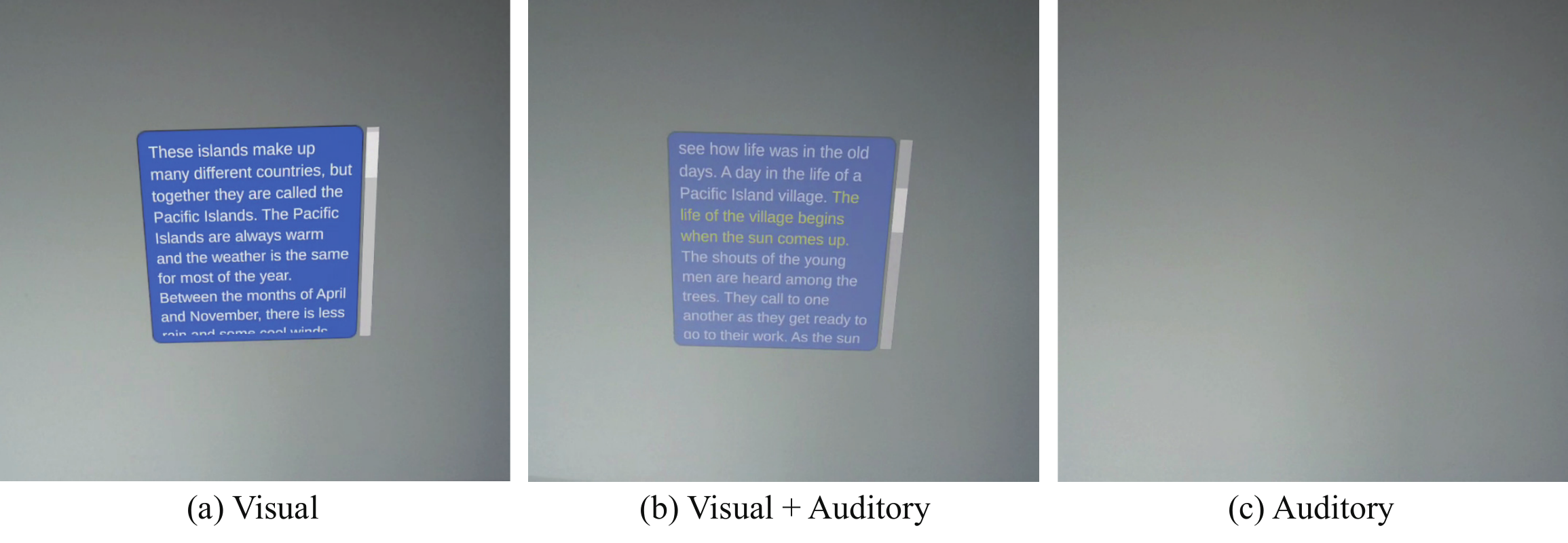}
    \vspace{-.4cm}
    \caption{Actual views of our prototype for each phase.}
    \vspace{-.4cm}
    \label{fig:user_view}
\end{figure*}

\subsection{Possible Interaction Scenarios}
\label{sec:interaction_scenario}
Switched Reading can be used in various scenarios where visual presentation of text information is temporarily unavailable or inappropriate. While this study does not address when or how to switch modalities, here we introduce several typical examples of interaction scenarios that take these practical considerations into account.

The most naive approach involves manual switching by the user (Figure~\ref{fig:interaction_scenario}a). The system can provide a simple toggle command, adopting various input methods such as the HWD's built-in interface controls, freehand gestures, or voice input.

More intelligent switching interactions can be designed by mapping modality changes to specific user behaviors or environmental contexts. For instance, as explored in the previous study using smartphones \cite{continuous-reading}, switching based on user activity, i.e., walking (auditory) versus stationary (visual) states, can support diverse scenarios and help prevent distracted walking (Figure~\ref{fig:interaction_scenario}b). In indoor settings like offices or libraries, modality switching based on seated (visual) versus standing (auditory) postures could also be beneficial.
Another strategy is to trigger switching by eye state, i.e., open (visual) versus closed (auditory) (Figure~\ref{fig:interaction_scenario}c). This would allow users to continue reading the text while intermittently closing their eyes, which may be beneficial in mitigating eye strain.

Furthermore, modality switching based on the environmental context is another promising strategy. For example, previous studies have often considered the placement of AR UIs based on the estimation of real-world surroundings and the possibility of user interaction with them \cite{Lindlbauer2019context,Li2025situationadapt}. Following this approach, the text presentation modality could adaptively switch depending on the presence of objects or people in the environment that are likely to interact with the user  (Figure~\ref{fig:interaction_scenario}d). This would help reduce the user's cognitive load and prioritize real-world awareness.

\subsection{Implementation}
\label{sec:implementation}
We implemented our prototype along with the two key interaction techniques described above. We used Meta Quest Pro\footnote{https://www.meta.com/quest/quest-pro/}, an HWD equipped with an eye tracker (per-eye resolution: 1832 x 1920 pixels, horizontal FOV: 106°, vertical FOV: 96°). The prototype was developed with a PC (Intel\textsuperscript{\textregistered} Core\texttrademark{} i9-9900K CPU @ 3.60 GHz, 32 GB RAM, GeForce RTX 4070Ti 16 GB) using Unity version 2022.3.3f1 and Oculus Integration, but it also works with the HWD alone without being tethered to the PC. For the speech synthesis engine, we used Microsoft Azure AI Speech\footnote{https://azure.microsoft.com/en-us/products/ai-services/ai-speech}.

The parameters of the text display interface (as shown in Figure \ref{fig:user_view} and Figure \ref{fig:experiment-VE}), such as the size and color of the virtual window, text color, and number of visible lines, were designed with reference to previous research and guidelines~\cite{vrReadingUIs,textReadability,reading-3dSurfaces,Hololens}, and they were optimized for viewing in Japanese. 
Specifically, the virtual window was placed 2.0 meters in front of the user, with its center positioned 5° below eye level. The window measured 1.0 m by 1.0 m, with 92\% opacity. Approximately nine lines of text were displayed within the window, each containing approximately 15 characters. The line spacing was set to 1.2 to balance the amount of visible text and the accuracy of gaze-based estimation of reading position. Text scrolling was controlled using the thumbstick on the Meta Quest Pro controller. In the following, we describe the implementation of the two key interaction techniques: (1) gaze-based voice playback and (2) correspondence-aware transition effect.


\subsubsection{Gaze-based Voice Playback}
\label{sec:gaze-based_playback}
To estimate the user's reading position on the displayed text, we adopted a gaze-based approach inspired by previous research~\cite{gazePrompt}. Our system uses the Meta Quest Pro's built-in eye tracker in this estimation process, which is illustrated in
Figure~\ref{fig:RPE}. First, for each character, the system pre-generates index data consisting of the sentence ID (sentence index in overall text) and phrase ID (delimiter index within each sentence separated by the reading point). During runtime, the system retrieves the user's gaze direction as a 3D vector in each frame. Since raw gaze data tend to be noisy, we applied a saccade-detection and smoothing algorithm from a previous work~\cite{smoothing} to filter the data. If the filtered gaze vector intersects with the text window, the character closest to the intersection point is identified through brute-force matching.
When the modality shifts from visual to auditory, the voice playback starts based on the estimated reading position (more detailed parameter tuning is described in Section ~\ref{sec:parameter_tuning_study}).

\subsubsection{Correspondence-aware Transition Effect}
\label{sec:transition_effect}
To support the user's contextual continuity in reading during modality switching, our interface introduces a transitional phase in which the visual and auditory modalities are active simultaneously (Figure~\ref{fig:System}). In the following, we describe the transition behaviors in detail.

\textbf{Visual $\to$ Auditory}
When switching from visual to auditory presentation, the interface transitions from fully opaque text display (Figure~\ref{fig:System}a\textcircled{\scriptsize 1}) to semi-transparent text with voice playback (Figure~\ref{fig:System}a\textcircled{\scriptsize 2}), thus providing a cue indicating that the modality is being changed. The voice playback starts based on the estimated reading position. During this transition, to support visual tracking, the currently spoken text (i.e., phrase divided by punctuation) is highlighted in yellow (Figure~\ref{fig:System}a\textcircled{\scriptsize 2}). A few seconds after the transition is initiated, the text display becomes fully transparent, and then only the audio modality continues (Figure~\ref{fig:System}\textcircled{\scriptsize 3}).

\textbf{Auditory $\to$ Visual}
While the audio is playing (Figure~\ref{fig:System}b\textcircled{\scriptsize 1}), the interface automatically sets the scroll position of the transparent text so that the spoken content stays at the top of a text window, although this window remains hidden from the user at this time. When the modality switch is triggered, the interface transitions to a semi-transparent display where the currently spoken sentence is highlighted and positioned at the top of the window (Figure~\ref{fig:System}b\textcircled{\scriptsize 2}). After a few seconds, the audio playback discontinues and the text becomes fully opaque to allow the user to continue visual reading (Figure~\ref{fig:System}b\textcircled{\scriptsize 3}).

\subsection{Parameter-tuning Study}
\label{sec:parameter_tuning_study}
\subsubsection{Overview}
We conducted a preliminary user study to fine-tune the key parameters of our interface. Eight students (six males, two females; mean age = 23.9 \(\pm\) 3.06) from our university or graduate school volunteered to participate in this study. Here, we examined two parameters that potentially influence the seamlessness of modality switching: (1) the starting point of voice playback---either at the beginning of the sentence containing the last fixated character (\textit{sentence-level}), at the beginning of the phrase divided by punctuation (\textit{phrase-level}), or from the fixated character itself (\textit{character-level})---and (2) the duration of the transition effect---1, 3, or 5 seconds.

Participants read a certain amount of text every 30 seconds under varied presentation modalities in a VR environment. First, they randomly experienced three conditions differing in the starting point of the voice playback and selected their preferred condition. Next, under their preferred voice playback condition, they randomly experienced three conditions differing in the duration of the transition effect and again selected their preferred condition. After completing these trials, we gathered qualitative feedback from the participants on their impressions and suggestions for potential improvement.

\subsubsection{Results and Discussion}
\textbf{Starting point of voice playback}:
Four participants preferred the phrase-level condition, while the other four favored the character-level condition. Those who chose the phrase-level playback noted that it started at a ``clean break'' and thus required less rereading. However, one participant also reported frustration when content they had already read visually was repeated audibly. In contrast, participants who preferred the character-level condition appreciated the minimal rereading required, but some found it disorienting when playback started at an unnatural point. Given that the preferences were evenly split, we believe this parameter should be customizable for each user in practical use. We also speculate that the phrase-level condition may be more effective when this interface is used in more distracting situations, such as while walking. In the subsequent user study, we decided to use phrase-level playback.

\textbf{Duration of transition effect}:
Three participants preferred a 3-second duration, while another three preferred 5 seconds. However, multiple participants stated that the most desirable duration differed depending on the \textit{direction} of the modality switch. When asked for clarification of this distinction, four participants indicated that a 3-second transition was ideal when switching from visual to auditory modalities. In contrast, three participants preferred a 1-second transition from auditory to visual, while four participants expressed a desire for the system to continue voice playback until the end of the sentence. Based on this feedback, in the subsequent user study we adopted a hybrid policy rather than setting a fixed duration: 3 seconds of transition effect for visual-to-auditory switching but continuation of the effect until the end of the currently spoken sentence for auditory-to-visual switching.

\section{User Study: Effects of Modality-Switching Techniques on Reading Experience}
\label{chap:userstudy}
We conducted a user study to evaluate the effectiveness of each of the two key interaction techniques in Switched Reading. We chose a fully immersive VR environment to minimize the influence of external factors and to clarify how different modality-switching methods affect the reading experience, rather than to establish their effectiveness in real-world AR/MR use. Specifically, we investigated the impact of the presence/absence of the gaze-based voice playback and the transition effect on user reading performance and user experience in a VR environment, where the text modality switches periodically. The experimental design of this study was approved by the Institute's Ethics Review Committee.

\subsection{Participants}
Sixteen undergraduate or graduate students (10 males, 6 females; mean age = 22.3 \(\pm\) 2.11) participated in the study. Participants with binocular visual acuity of 0.7 or better, including those with contact lenses, were recruited (we requested in advance that they refrain from participating in the experiment while wearing eyeglasses due to the potential for eye tracking inaccuracies). None of the participants had prior knowledge of this study. Seven participants used an HWD for the first time, and the remaining nine had used one at least twice. Eleven participants reported using audio-based media (e.g., radio, podcasts, audiobooks) at least once a month.

\begin{figure*}[t]
    \centering
    \raisebox{2.5\baselineskip}{%
    \begin{minipage}[t]{0.54\textwidth}
        \centering
        \begin{subfigure}[t]{0.57\linewidth}
            \centering
            \includegraphics[width=\linewidth]{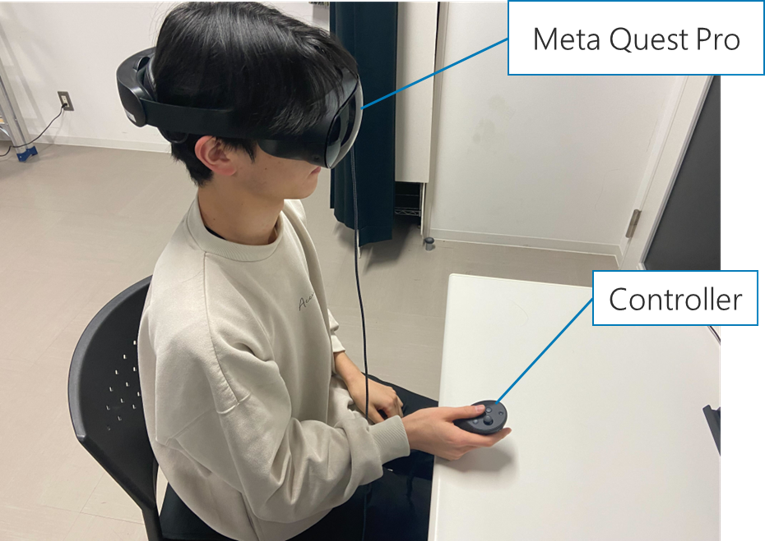}
            \subcaption{Experimental setup.}
            \label{fig:experimental-setup}
        \end{subfigure}
        \hfill
        \begin{subfigure}[t]{0.41\linewidth}
            \centering
            \includegraphics[width=\linewidth]{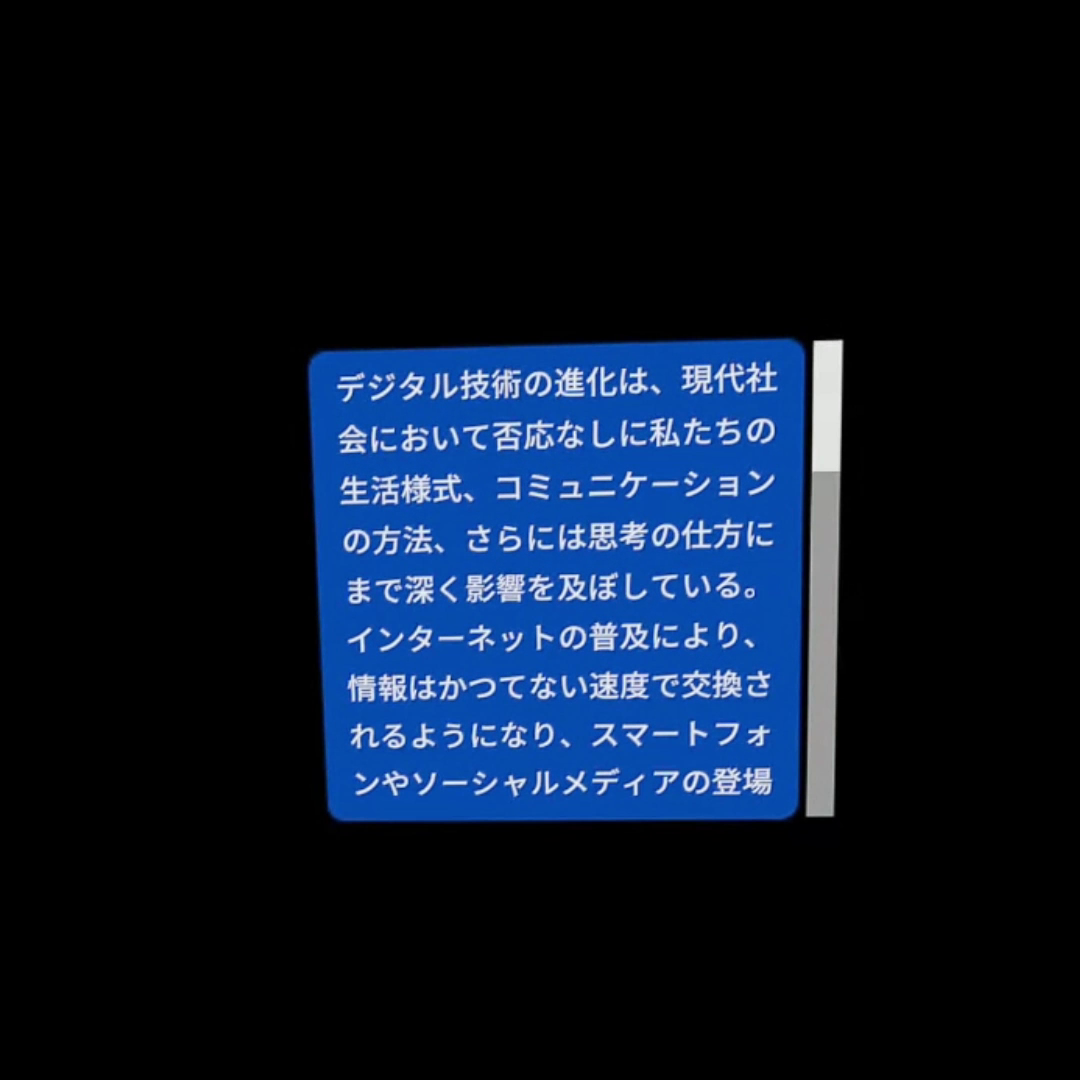}
            \subcaption{Virtual environment presented to participants.}
            \label{fig:experiment-VE}
        \end{subfigure}
        \caption{Apparatus.}
        \vspace{-.4cm}
        \label{fig:apparatus}
    \end{minipage}%
    }
    \hfill
    \begin{minipage}[t]{0.44\textwidth}
        \centering
        \includegraphics[width=0.9\linewidth]{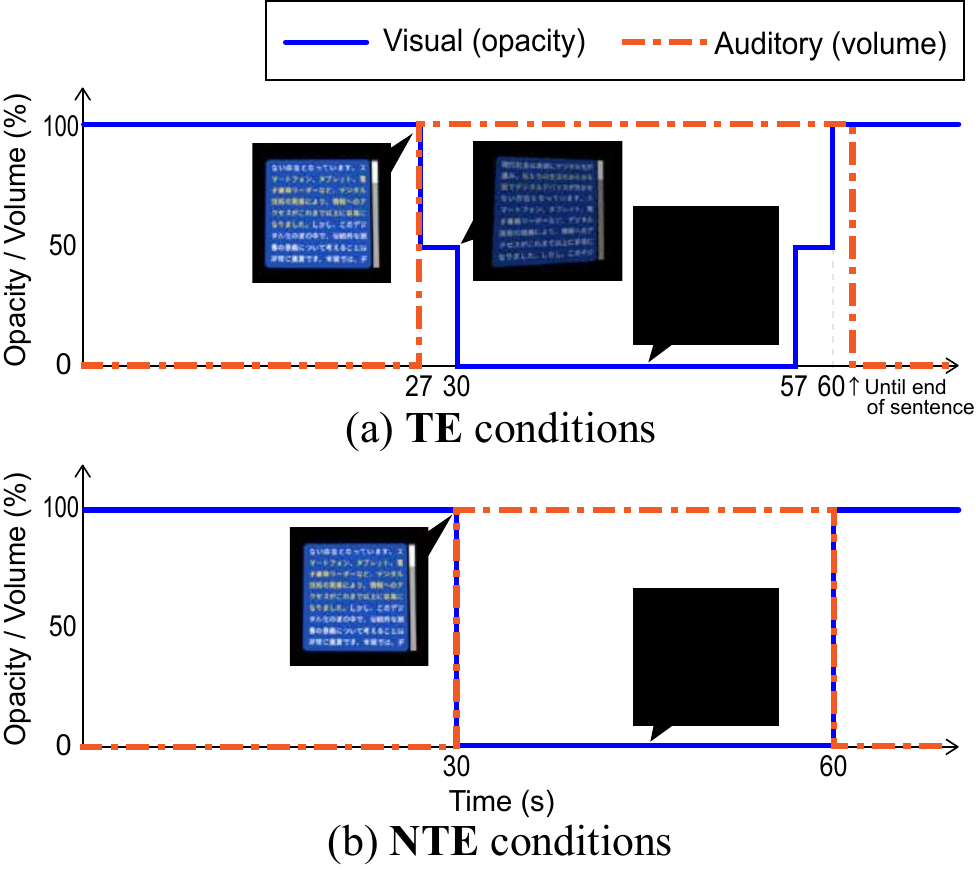}
        \captionof{figure}{Visual-auditory differences between TE and NTE conditions.}
        \vspace{-.4cm}
        \label{fig:experiment-task}
    \end{minipage}
\end{figure*}

\subsection{Experimental Design}
The study followed a within-subjects design with two independent variables: \textit{voice playback} (gaze-based voice playback (\textbf{GVP}) vs. scroll-position-based voice playback (\textbf{SVP})) and \textit{transition effect} (with transition effect (\textbf{TE}) vs. no transition effect (\textbf{NTE})). For voice playback, we designed SVP as a baseline condition based on the previous study of mobile continuous reading~\cite{continuous-reading}, in which voice playback starts from the phrase shown at the top of the currently displayed screen. As for the baseline of transition effect, we implemented NTE where visual and auditory presentations were presented sequentially without any temporal overlap.

In the task, participants read a specific text for five minutes while the modality switched between visual and auditory every 30 seconds. This setup simulates scenarios where the appropriate modality changes over time (e.g., walking through a city and occasionally stopping at intersections with ``Walk''/``Wait'' signals) while minimizing the influence of external factors. We chose a 30-second interval to collect enough switching events without disrupting the task due to loss of textual context. Participants were instructed to read the text both for comprehension and with the goal of reading as quickly as possible within a comfortable range.

The dependent variables included comprehension accuracy, reading speed (characters per minute (CPM)), number of visual rereads, visual-auditory offset, NASA-TLX \cite{hart1988development}, and System Usability Scale (SUS) \cite{bangor2008empirical}.

For reading speed, we used characters per minute (CPM) rather than words per minute (WPM) because the stimuli were Japanese, where reading/listening pace is closely tied to phonological units (morae).
For the number of rereads, we counted them based on participants' gaze behavior: We determined the most frequently gazed-at line number within each time window and considered a reread to have occurred when the gaze returned by more than three lines compared with the previous window. We also implemented an auditory reread function using the controller's thumbstick, but we did not analyze it because it was rarely used in our pilot test.

Visual-auditory offset was calculated as the character difference between the last fixated character and the first character of the voice playback. We assumed that this metric indicated how much the user's (estimated) reading position shifted when switching from visual to auditory modality, thus reflecting how smoothly the user switched modalities.

Subjective metrics included NASA-TLX and SUS, which were rated on a 7-point Likert scale. We also obtained qualitative comments on the user experience during the visual-auditory transition and overall text reading comprehension.

\subsection{Apparatus}
Figures~\ref{fig:experimental-setup} and ~\ref{fig:experiment-VE} show an overview of the experimental setup and the virtual environment presented to the participants, respectively. We used the same system described in Section \ref{sec:implementation} with one modification: All synthesized voice was pre-generated and stored locally. In our pilot test, we found that our system took approximately 1.3 seconds to initiate speech synthesis, which could significantly affect the seamlessness of the modality switching experience. To investigate the user experience without such implementation-related limitations, we virtually eliminated the playback delay by preloading audio files locally (starting in less than $10^{-5}$ seconds).

The text materials and comprehension questions were generated using GPT-4o, a large language model (LLM) by OpenAI\footnote{https://openai.com/index/hello-gpt-4o/}. The input prompt specified the target character count (around 7,000 characters) and genre (novel, essay, or commentary on sports/arts). We then slightly edited the generated texts manually to improve readability. The difficulty level of all text materials used was rated as ``\textit{low-intermediate}'' (i.e., normal difficulty, the fourth highest of six levels) by jReadability\footnote{https://jreadability.net/sys/}, a system for assessing the difficulty of text in Japanese.
Each text was accompanied by three multiple-choice comprehension questions created by the LLM, which were further refined by the authors (e.g., ``\textit{Where did Akira and his friends first go stargazing? a. Seaside, b. Mountain peak, c. Downtown, d. Forest}''). These questions were designed to assess whether participants correctly understood and remembered the content of the text. The content and difficulty level we set were confirmed to be appropriate in our informal pilot test.

\subsection{Procedure}
\subsubsection{Overall Procedure}


The experiment involved four conditions combining two voice playback conditions (GVP and SVP) and two transition effect conditions (TE and NTE). The order of the four conditions and the four texts used in the trials was counterbalanced across participants. The entire experiment took approximately 120 minutes per participant, including short breaks between conditions. 

Here, we describe this procedure in detail. First, the experimenter explained the task and obtained informed consent. The participant then put on the HWD, and inter-pupillary distance was calibrated. After this preparation, the participant completed a practice trial. In this trial, participants were introduced to the interface used in the main trials, their eye tracking was calibrated using a function of Quest Pro, and they were given a practice text of similar difficulty to the main tasks. They received instructions on how to operate the interface and were then asked to read the entire text. Afterward, they answered three multiple-choice questions to assess their text comprehension. They practiced using the interface until they felt comfortable with it.

Following a short break, participants proceeded to the main trials. They completed the main reading task with each interface and then answered three multiple-choice comprehension questions. After completing the trials for each interface, participants filled out a Google Form questionnaire assessing NASA-TLX and SUS. Then, we conducted a semi-structured interview to mainly explore how the conditions influenced the user experience during the visual-auditory transition as well as overall text reading comprehension.

\subsubsection{Task Procedure}
Here, we explain the procedure used for the reading task. First, participants calibrated the text display location after assuming a comfortable reading posture and pressing a button on the controller. During the reading task, the text transitioned every 30 seconds between visual and auditory modalities. At the start of each task, the text was always initially visible. In the TE conditions, the text window changed its opacity three seconds before switching the modality. The task ended when the participant either finished reading the entire text and pressed a button on the controller or when the final sentence of the audio was completed. Although a five-minute time limit was set for the task, all participants were able to complete it within that time. Task duration, used to calculate reading speed, was measured from the time the initial text was displayed to when the task was completed.

\subsection{Results}
All participants successfully completed all trials without problems. Since Shapiro-Wilk tests showed most data distributions did not satisfy normality, we used non-parametric statistical methods for analysis. Note that \textbf{GVP-TE} is the condition corresponding to our proposed interaction technique (i.e., the combination of the two interaction techniques within the Switched Reading framework).

\subsubsection{Objective Measures}
\begin{figure*}[t]
    \centering
    \begin{minipage}[t]{0.32\textwidth}
        \centering
        \includegraphics[width=\textwidth]{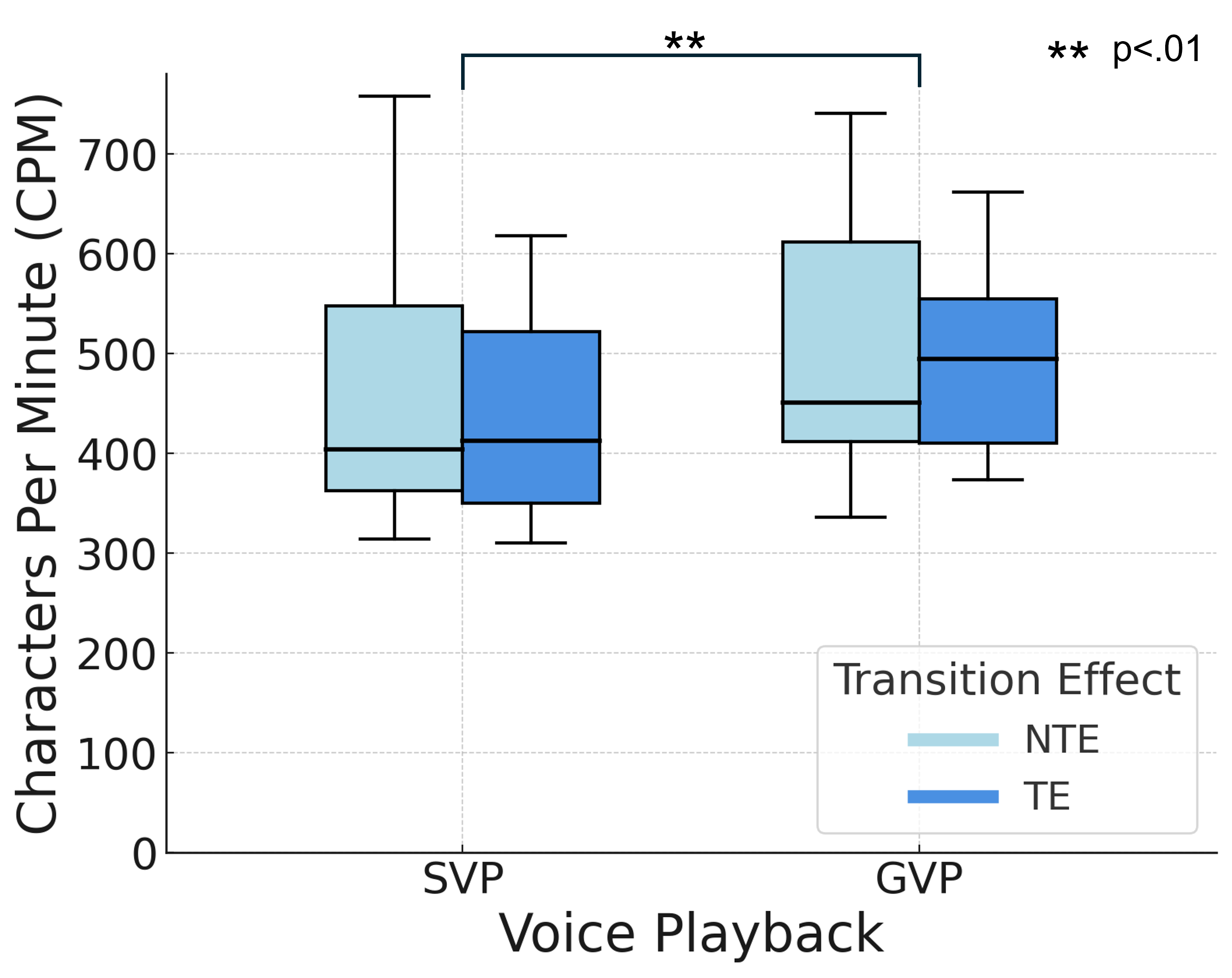}
        \vspace{-.4cm}
        \caption{Reading speed (CPM)}
        \label{fig:cpm2}
    \end{minipage}
    \hfill
    \begin{minipage}[t]{0.32\textwidth}
        \centering
        \includegraphics[width=\textwidth]{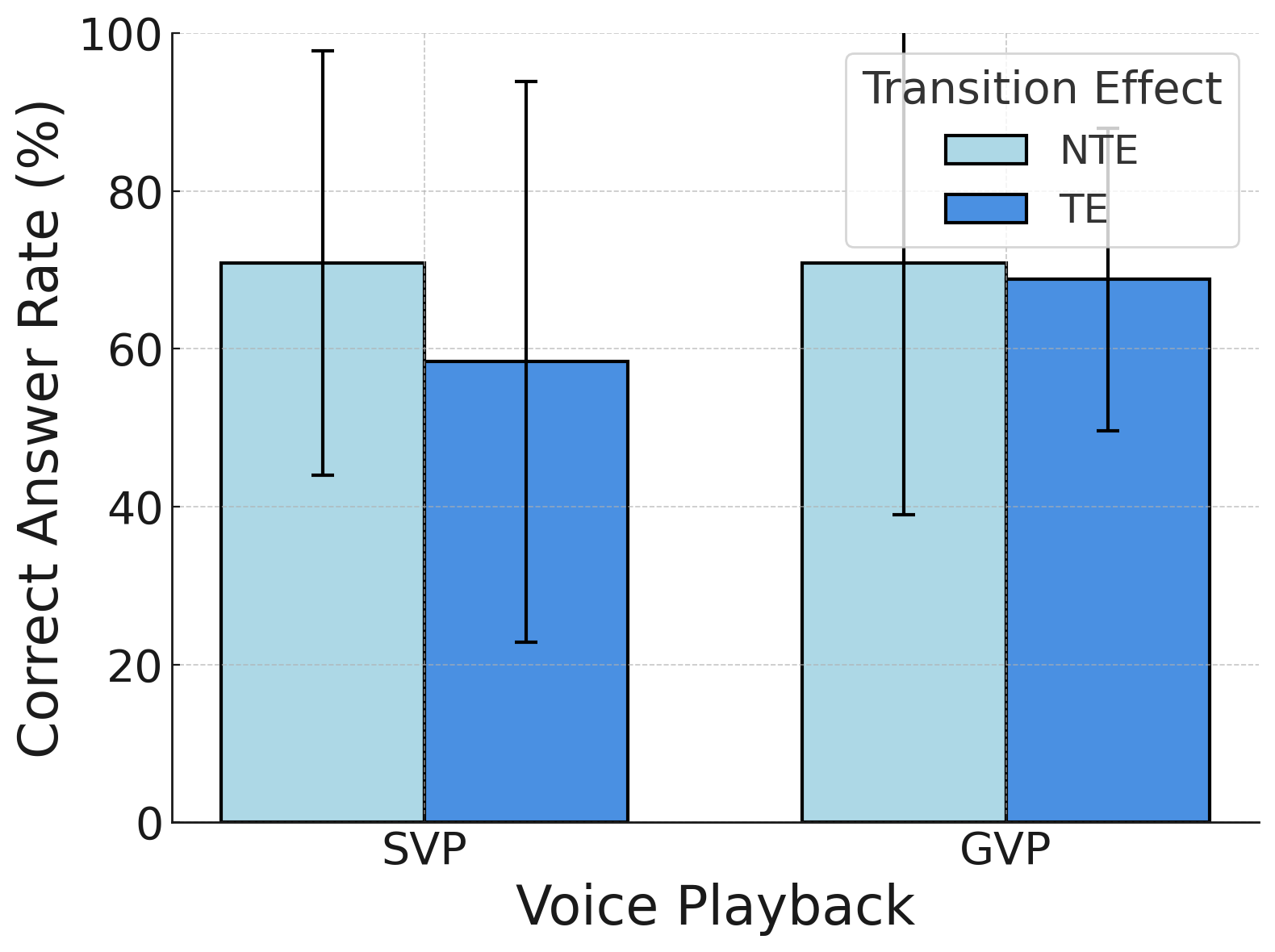}
        \vspace{-.4cm}
        \caption{Comprehension accuracy}
        \label{fig:correct2}
    \end{minipage}
    \hfill
    \begin{minipage}[t]{0.32\textwidth}
        \centering
        \includegraphics[width=\textwidth]{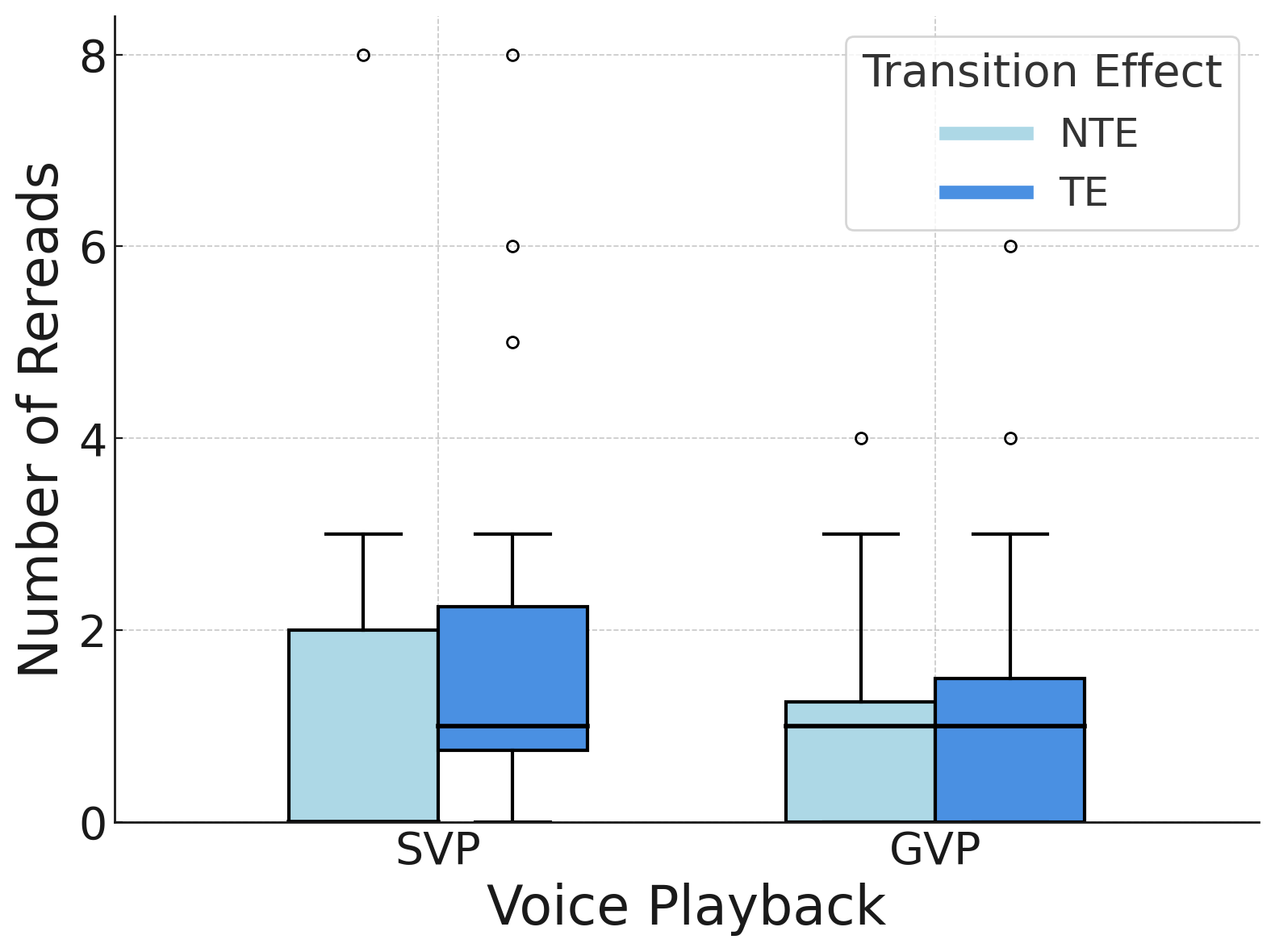}
        \vspace{-.4cm}
        \caption{Number of rereads (Visual)}
        \label{fig:readingBack2}
    \end{minipage}
\end{figure*}

\begin{figure*}[t]
    \centering
    \begin{minipage}[t]{0.32\textwidth}
        \centering
        \includegraphics[width=\textwidth]{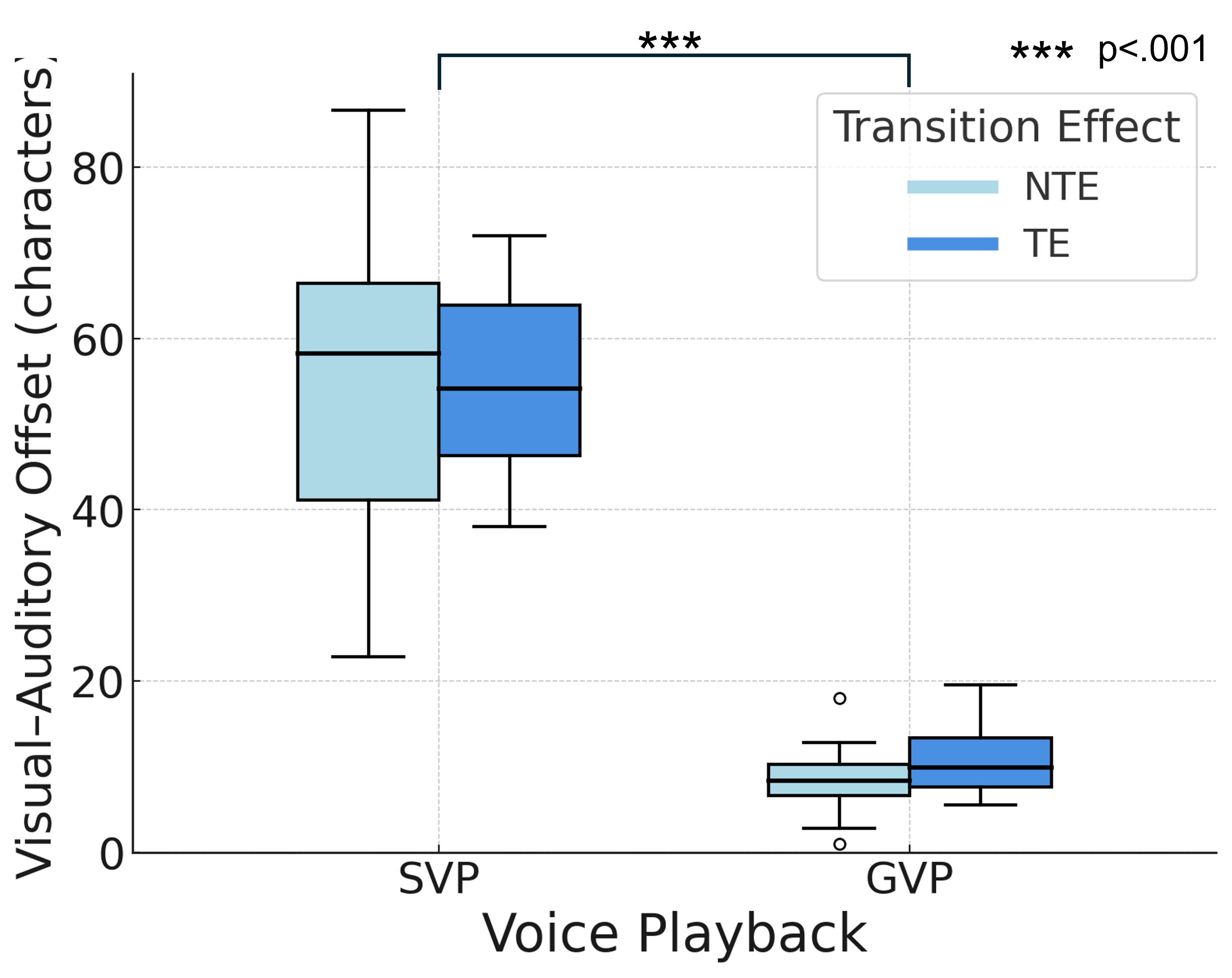}
        \vspace{-.4cm}
        \caption{Visual-auditory offset}
        \vspace{-.4cm}
        \label{fig:difference2}
    \end{minipage}
    \hfill
    \begin{minipage}[t]{0.32\textwidth}
        \centering
        \includegraphics[width=\textwidth]{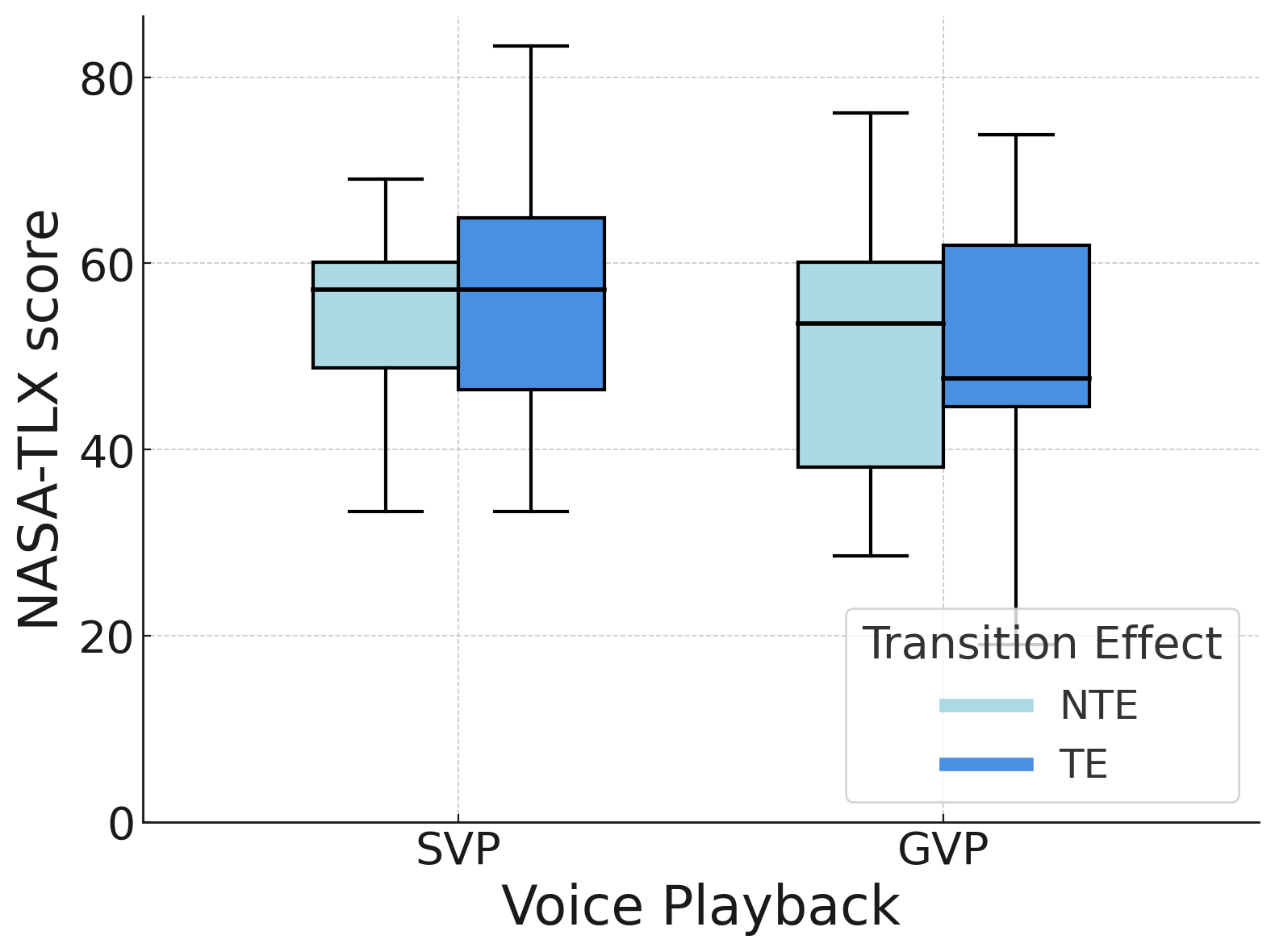}
        \vspace{-.4cm}
        \caption{NASA-TLX score}
        \vspace{-.4cm}
        \label{fig:nasatlx2}
    \end{minipage}
    \hfill
    \begin{minipage}[t]{0.32\textwidth}
        \centering
        \includegraphics[width=\textwidth]{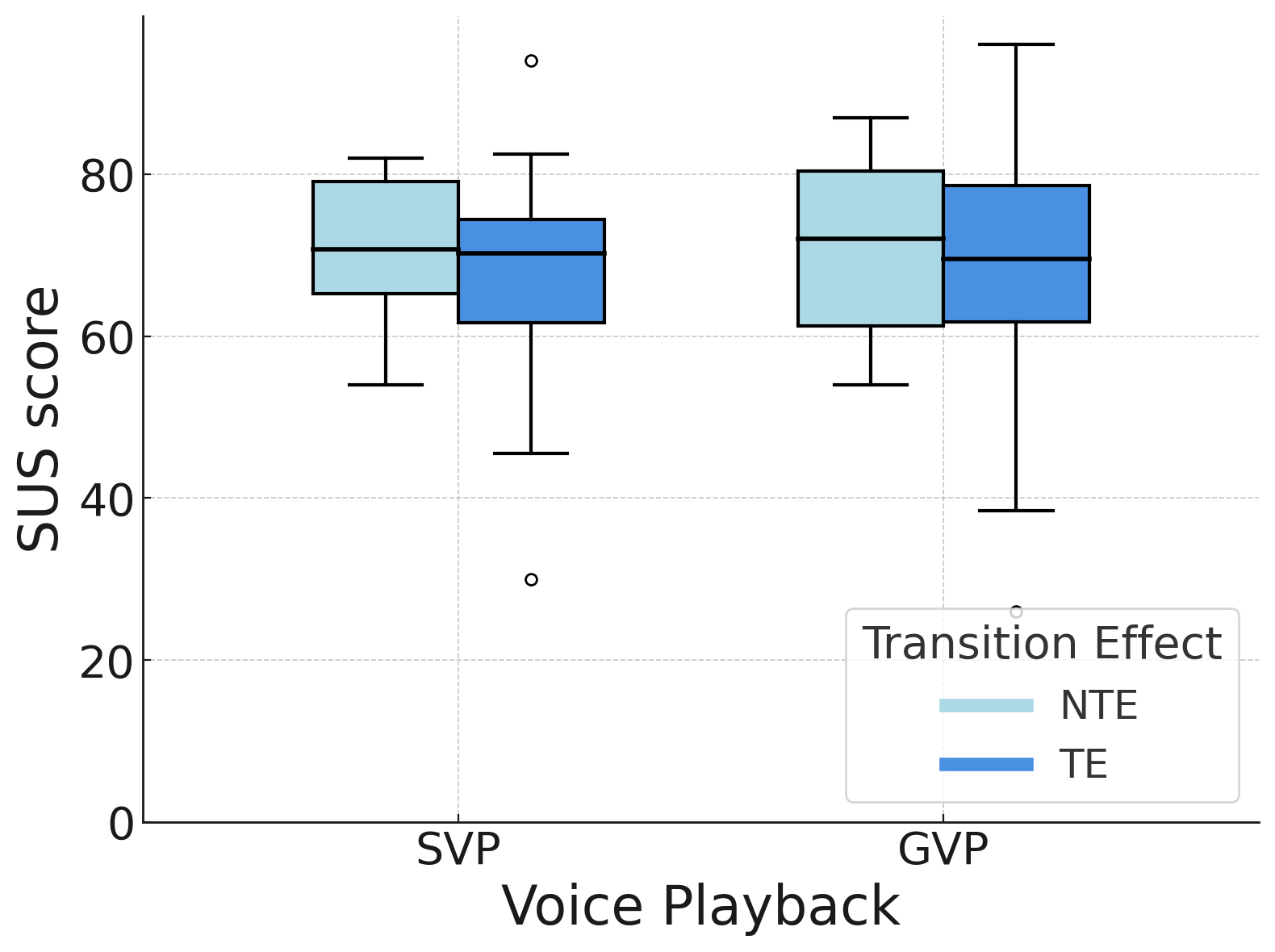}
        \vspace{-.4cm}
        \caption{SUS score}
        \vspace{-.4cm}
        \label{fig:sus2}
    \end{minipage}
\end{figure*}

\noindent\textbf{Reading Speed (CPM)}.
Figure~\ref{fig:cpm2} shows CPM across conditions. A two-way Aligned-Rank Transform (ART) ANOVA~\cite{wobbrock2011aligned} showed showed no significant interaction between voice playback and transition effect ($F(1, 15) = 0.10, p = 0.752, \eta_p^2 = 0.007$). 
Regarding the main effect, we found a significant main effect of voice playback ($F(1, 15) = 21.21, p < .001, \eta_p^2 = .586$). The Wilcoxon signed-rank test showed that GVP yielded significantly higher CPM than did SVP ($r = -0.789$, $W = 7.0$, $p = 0.002$).
In contrast, there was no significant main effect of transition effect ($F(1, 15) = 1.94, p = 0.184, \eta_p^2 = 0.115$).

\noindent\textbf{Comprehension Accuracy}.
Figure~\ref{fig:correct2} displays comprehension accuracy across conditions. ART ANOVA showed no significant interaction (interaction: $F(1, 15) = 0.24, p = 0.628, \eta_p^2 = .016$) or main effect (voice playback: $F(1, 15) = 0.25, p = 0.625, \eta_p^2 = 0.016$; transition effect: $F(1, 15) = 1.67, p = 0.216, \eta_p^2 = 0.100$).

\noindent\textbf{Number of Rereads}.
Figure~\ref{fig:readingBack2} shows the number of visual rereads. ART ANOVA indicated no significant interaction (interaction: $F(1, 15) = 2.04, p = 0.174, \eta_p^2 = 0.120$) or main effect (voice playback: $F(1, 15) = 0.53, p = 0.479, \eta_p^2 = 0.034$; transition effect: $F(1, 15) = 1.64, p = 0.220, \eta_p^2 = 0.099$).

\noindent\textbf{Visual-Auditory Offset}.
Figure~\ref{fig:difference2} shows the visual-auditory offset by number of characters. ART ANOVA showed no significant interaction effect ($F(1, 15) = 0.20, p = 0.659, \eta_p^2 = 0.013$). However, a significant main effect of voice playback was found ($F(1, 15) = 65.08, p < 0.001, \eta_p^2 = 0.813$). The Wilcoxon signed-rank test revealed that the offset was significantly smaller with GVP than with SVP ($r = 0.879$, $W = 136.0$, $p < 0.001$). No significant main effect was found for transition effect ($F(1, 15) = 0.30, p = 0.590, \eta_p^2 = 0.020$).

\subsubsection{Subjective Measures}

\noindent\textbf{NASA-TLX}.
Figure~\ref{fig:nasatlx2} shows average NASA-TLX scores across the six subscales for each condition. Out of the four conditions, our proposed combination (i.e., GVP-TE) yielded the lowest average score ($M = 49.85, SD = 16.11$). However, ART ANOVA revealed no significant interaction (interaction: $F(1, 15) = 3.01, p = 0.103, \eta_p^2 = 0.167$) or main effect (voice playback: $F(1, 15) = 2.46, p = 0.138, \eta_p^2 = 0.141$; transition effect: $F(1, 15) = 0.02, p = 0.902, \eta_p^2 = 0.001$).

\noindent\textbf{SUS}.
Figure~\ref{fig:sus2} gives the SUS scores. ART ANOVA showed no significant interaction ($F(1, 15) = 0.09, p = 0.775, \eta_p^2 = 0.006$) or main effect (voice playback: $F(1, 15) = 0.14, p = 0.711, \eta_p^2 = 0.009$; transition effect: $F(1, 15) = 0.27, p = 0.608, \eta_p^2 = 0.018$).



\begin{figure}[t]
    \centering
    \includegraphics[width=0.8\columnwidth]{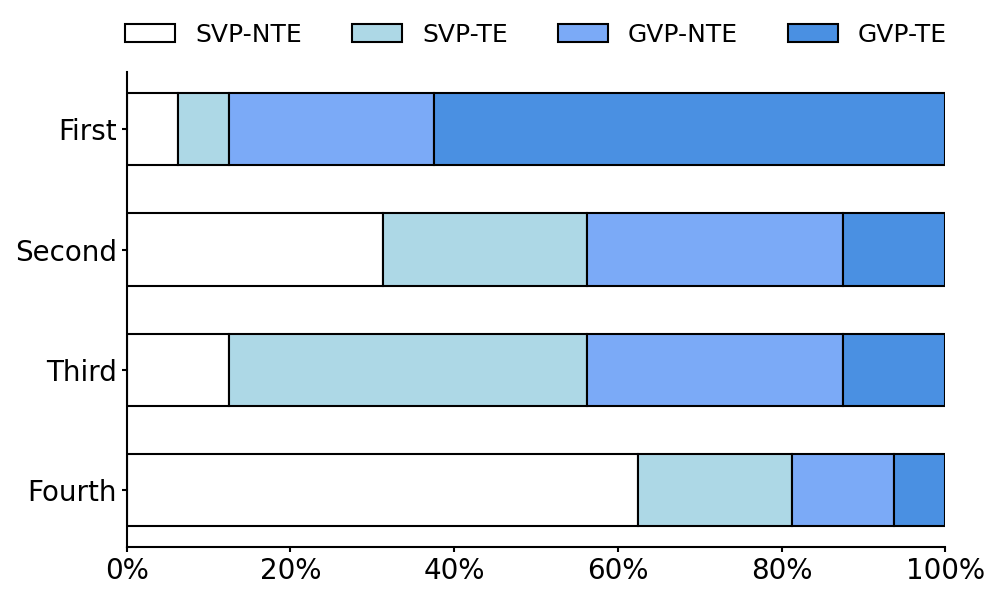}
    \caption{Participant preferences for the four conditions.}
    \vspace{-.4cm}
    \label{fig:preference2}
\end{figure}

\noindent\textbf{Preferences and Comments}.
Figure \ref{fig:preference2} illustrates the participants' preferences for the four conditions in order of favorability. Among the 16 participants, 10 selected GVP-TE as their most preferred condition, 4 chose GVP-NTE, 1 selected SVP-TE, and 1 chose SVP-NTE. The reasons cited for preferring GVP-TE included the ability to seamlessly switch between visual and auditory modalities (P2, P4, P6, P7, P9, P14, P16), ease of understanding the text (P7, P13, P16), enhanced focus during reading (P5, P6), and the ability to maintain an awareness of their reading position (P13, P14).

When asked about the audio playback's starting position (GVP vs. SVP), participants overwhelmingly expressed positive opinions about GVP. This condition was praised for achieving faster reading (P1, P3, P12), better comprehension of the text (P1, P5), a more natural experience of reading/hearing text (P12), and the participant's ability to focus more on reading (P6). 

In contrast, the SVP condition received predominantly negative feedback. Participants noted that they had to constantly monitor the scroll position (P5, P6, P14, P15) and sometimes lost track of their reading position (P2). 
Additionally, several participants (P7, P9, P10, P13) mentioned that during the visual-to-auditory transition, voice playback often started from a section they had already read, which they found repetitive and annoying. 
On the other hand, some participants stated the opposite: While the same section was repeated, it caused minimal confusion (P8), or it was found preferable (P11) or even helpful for better comprehension (P3, P11, P16).


Regarding the transition effect (TE vs. NTE), most participants expressed a more favorable impression of the TE condition. Interestingly, many participants (P1, P2, P4, P5, P6, P13, P15, P16) mentioned that they could better comprehend the text under TE compared to NTE. Other positive feedback included observations that the text was easier to read (P5, P7, P13, P14), the transitions were smoother (P15, P16), it allowed them to mentally prepare for the switch (P11), they did not lose track of their reading position (P14), and they could focus better on reading (P6). In contrast, some participants mentioned a gap between their reading speed and the voice playback speed, which they found bothersome (P1, P3), and that their reading pace slowed down (P1). Additionally, one participant (P10) preferred the NTE condition, citing its simplicity and fewer changes in the presentation format.

\subsection{Discussion}
\subsubsection{Voice Playback}
One of the most notable findings was that \textbf{the visual-auditory offset was significantly smaller in the GVP conditions than in the SVP conditions}. While this result may seem intuitive, it is worth highlighting that the voice playback position in SVP was more than 50 characters away from the gaze position during modality switches, whereas in GVP, the offset was reduced to approximately 10 characters. Playback method also had large effects on reading speed ($\eta_p^2=.586$, $|r|=.789$), although their magnitude may not generalize to naturalistic AR/MR settings. Interview feedback also suggests that GVP facilitated faster reading. Moreover, while we hypothesized that GVP would result in fewer rereads, no significant difference was observed between conditions. This is perhaps because the participants were focused enough on the reading task that there was little rereading, regardless of the condition. Results may differ in more complex tasks, such as walking while avoiding obstacles, where reading performance is known to decline significantly \cite{klose2019text}. 

In contrast, SVP was less preferred by participants. From the interviews, we speculated that the primary reason was the perception of participants that they were compelled to adapt their reading behavior to the system. Specifically, they felt they had to constantly monitor the scroll position in anticipation of modality switches.

Despite these differences, SVP did not differ significantly from GVP in terms of comprehension accuracy. This might be because SVP was designed to always start voice playback from the beginning of the visible text window, resulting in a backward offset (i.e., voice playback started from a section the user had already read). While some participants found this repetitive and annoying---and it actually led to a decrease in reading speed---others noted that it helped reinforce their understanding. In our pilot test, we also considered another design in which audio playback starts from the middle of the text window in the SVP condition. However, this design should be avoided because it risks starting playback from sections the user had not yet read.

\subsubsection{Transition Effect}
Although several participants perceived TE conditions as helpful for understanding the text, comprehension accuracy did not differ significantly between TE and NTE. We therefore interpret these comments as perceived support for maintaining textual context, rather than as evidence of improved comprehension performance. 

Furthermore, the TE conditions did not significantly affect reading speed. We interpret this as the transition effect functioning as a ``safety net'' to prevent users from losing their reading position, thus providing a greater sense of control without directly improving performance. More specifically, the NTE conditions may not have caused substantial problems in completing the tasks because the reading position was accurately estimated (in GVP conditions) or the voice playback position was always fixed (in SVP conditions). Again, participants in this study were able to focus on the reading task, but the transition effect may be more useful in more distracting situations that also require attention to the real world.

Furthermore, the lack of significant advantages of the TE conditions in subjective metrics, such as NASA-TLX and SUS, might be attributed to certain design shortcomings. First, participants were bothered by the mismatch between their visual reading speed and the voice playback speed. Addressing this may require features that allow users to customize the playback speed or dynamically adjust it based on gaze position. Second, some participants (P5, P9) felt that the 3-second duration of the transition effect was too short. While this duration was determined based on our preliminary test, it may have imposed time pressure on users to achieve the transition within a limited time frame. To address this, the transition duration and profile could be customized or adapted to the switching direction, reading speed, and context, while persisting until the user has successfully tracked the audio playback position with their gaze.

\begin{table*}[th]
    \centering
    \caption{Results of application study with statistic values of Wilcoxon signed-rank tests.}
    \begin{tabular}{ccccc}
        \hline
        Metric & Manual Switch & Activity-based Switch & $r$ & $p$\\
        \hline
        SUS score & 79.4 ($SD=9.54$) & 66.0 ($SD=22.4$) & 0.477 & 0.067 \\
        Number of modality switches & 4.75 ($SD=3.93$) & 10.75 ($SD=0.750$) & 0.069 & 0.004**\\
        Audio/Visual duration used & 24.4 \%/75.6 \% & 50.3 \%/49.7 \% & - & -\\ 
        Reading speed (CPM) & 490 ($SD=222$) & 423 ($SD=126$) & 0.918 & 0.110 \\
        Walking speed (m/s) & 0.833 ($SD=0.168$) & 0.952 ($SD=0.160$) & 0.821 & 0.003** \\
        \hline
    \end{tabular}
    \vspace{-.4cm}
    \label{tab:applicationstudy}
\end{table*}
\section{Application Study}
\subsection{Overview}

To derive more practical design guidelines of Switched Reading, we built an application designed for reading while walking and investigated its user experience by obtaining objective/subjective data. 
Twelve volunteers (10 males, 2 females, and no others) from a local university participated in the study. Their mean age was 22.67 years ($SD = 1.18$). We used equipment similar to that in the previous user study described in Section~4. This application study was covered by the same institutional ethics approval as the previous user study.

The application implemented two switching methods corresponding to those shown in Figure~\ref{fig:interaction_scenario} a and b: \textit{\textbf{Manual Switch}} and \textit{\textbf{Activity-based Switch}}.
For the Manual Switch, we simply used button presses on the Meta Quest Pro controller. The Activity-based Switch automatically switched between visual and auditory modalities based on the user's stationary or walking state. To detect these states, we calculated the walking speed by acquiring the moving average of the HMD's positional coordinates at approximately 70 Hz. The state transitions between stationary and walking were triggered when the walking speed remained below or above a predetermined threshold for a certain duration.

\begin{figure}[t]
    \centering
    \includegraphics[width=0.6\linewidth]{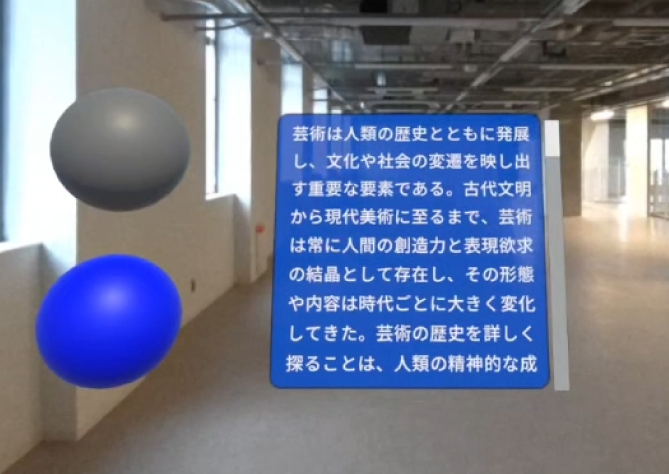}
    \caption{Participant view in application study.}
    \vspace{-.4cm}
    \label{fig:applicationstudy}
\end{figure}

The task was designed as a dual-task scenario that combined walking and text reading, following previous studies using similar protocols \cite{klose2019text,lu2020glanceable,reading-onSmartGlass}. For the walking task, participants walked back and forth along a 1.5 m wide and 25 m long empty corridor. Additionally, as shown in Figure \ref{fig:applicationstudy}, two virtual spheres simulating traffic signals were displayed at fixed positions in their visual field, switching between red and blue colors at 30-second intervals with random fluctuations of a few seconds. Participants were instructed to walk at a safe, comfortable pace when the signal was blue and to remain stationary when it was red. For the reading task, participants were instructed to read the text (as in the previous user study) as much as possible while maintaining comprehension throughout the trial. For the Manual Switch, participants were instructed that they could switch modalities at any time during the trial. After experiencing each of the two modality-switching methods, participants completed the SUS questionnaire and participated in a semi-structured interview. The presentation order of the two methods was counterbalanced across participants. Metrics applied can be found in Table~\ref{tab:applicationstudy}.

\subsection{Results and Discussion}
Table~\ref{tab:applicationstudy} presents the results of SUS scores and other objective metrics, as well as the results of Wilcoxon signed-rank tests comparing the two modality-switching methods.


Regarding the SUS scores, the Manual Switch method provided higher mean scores than the Activity-based Switch (without significant difference). Consistently, the majority of participants provided positive feedback for the Manual Switch, with many emphasizing the advantage of switching modalities at arbitrary times (N=5). A minority of participants mentioned the tediousness (N=2) or low necessity (N=2) of switching. In contrast, the Activity-based Switch received mixed feedback. Positive comments highlighted that automatic switching was effortless (N=2) or reasonable (N=1), and participants appreciated the naturalness (N=2) and enhanced focus (N=1) of the reading experience. Conversely, negative comments pointed out the lack of user control over modality selection (N=2) and the disruption of context during transitions (N=2).

Regarding participants' modality switching behavior, as shown in Table~\ref{tab:applicationstudy}, the Manual Switch method resulted in fewer switches and longer periods of visual presentation compared to the Activity-based Switch method. Indeed, many participants showed a preference for visual presentation. Some participants (N=3) mentioned that the ability to use only visual modality was an advantage of Manual Switch, while others (N=3) noted that difficulty in acquiring information through auditory presentation was a drawback of Activity-based Switch. This preference may partly reflect the empty and visually uncomplicated corridor, where participants perceived relatively little need to switch away from visual reading. Based on these observations, we recommend that \textbf{modality switching be customizable by users as a hybrid approach combining both automatic and manual switching}, rather than being fully automatic. Furthermore, when designing automatic switching, \textbf{visual presentation should be prioritized as long as safety conditions permit}.

In contrast, the Activity-based Switch provided a 13.7\% slower reading speed but a 14.3\% faster walking speed than Manual Switch. The slower reading speed was likely due to the longer duration of auditory presentation in the Activity-based Switch. The faster walking speed, as intended, suggests that \textbf{automatic switching based on walking activity may help prevent distracted walking and enhance perceived safety during locomotion}. In fact, one participant explicitly appreciated the unobstructed view of the real environment provided during walking. Since this study utilized a straight corridor with minimal obstacles or pedestrians, the advantages of activity-based switching may be more pronounced in real-world situations requiring greater awareness of the environment. Future investigation in such contexts is necessary.

\section{Conclusion, Limitations, and Future Work}
We proposed Switched Reading and investigated gaze-based voice playback and a correspondence-aware transition effect through a controlled user study and an application study. These studies yielded the following design implications for the proposed interface and possibly similar cross-modality reading systems:
\begin{itemize}
    \item \textbf{Gaze-based voice playback is recommended}, since users generally preferred it and read faster with fewer gaze shifts during modality switching. Optionally replaying sections already read visually may support comprehension, but the system must avoid starting from unread sections.
    \item \textbf{The transition effect may be beneficial}, since users generally preferred it and perceived support for comprehension, although it did not objectively improve reading speed or comprehension accuracy. Further improvements should reduce the mismatch between visual reading and voice playback speeds and maintain the effect until users have mapped the correspondence between modalities.
    \item \textbf{Visual presentation should be prioritized whenever safety conditions permit}, since visual reading is faster and was more favored by users in our application study. However, considering that some users may prefer auditory presentation and irregular context changes may occur, it is advisable to ensure that manual switching options are always available and that users can fine-tune the switching triggers to suit their habits.

\end{itemize}

Nevertheless, our studies were conducted in relatively controlled settings, with predictable 30-second switching intervals and few external interruptions. Therefore, the effects of frequent or unexpected switching in more complex environments remain unclear. Gaze-estimation errors may also cause playback or highlighting to begin at unintended positions, particularly during locomotion. Furthermore, participants differed in their modality and transition preferences, suggesting that practical systems should support personalization and manual overrides rather than treating the tested settings as universally optimal.


In addition, this study was conducted using only Japanese text. While the fundamental design of the proposed interface is likely applicable to other languages, differences in punctuation frequency and the presence/absence of spaces between words may require subtle variations in the design of the transition effect.

Finally, our prototype focused mainly on plain text. In reality, books and academic papers often include figures and tables in addition to text. In future work, to make such visual elements accessible in the auditory modality, we plan to explore methods such as converting them into descriptive text using large language models.


\acknowledgments{
This work was supported by JSPS KAKENHI (JP25K03160). Generative AI tools, including ChatGPT, were used to create text materials for the user study and assist in translating the manuscript from the authors' local language into English; the final English was checked by a human editor.
}

\bibliographystyle{abbrv-doi}

\bibliography{reference}

@inproceedings{reading-onthego,
author = {Vadas, Kristin and Patel, Nirmal and Lyons, Kent and Starner, Thad and Jacko, Julie},
title = {Reading on-the-go: a comparison of audio and hand-held displays},
year = {2006},
isbn = {1595933905},
publisher = {Association for Computing Machinery},
address = {New York, NY, USA},
url = {https://doi.org/10.1145/1152215.1152262},
doi = {10.1145/1152215.1152262},
booktitle = {Proceedings of the 8th Conference on Human-Computer Interaction with Mobile Devices and Services},
pages = {219–226},
numpages = {8},
location = {Helsinki, Finland},
series = {MobileHCI '06}
}

@inproceedings{chua2016positioning,
  title={Positioning glass: Investigating display positions of monocular optical see-through head-mounted display},
  author={Chua, Soon Hau and Perrault, Simon T and Matthies, Denys JC and Zhao, Shengdong},
  booktitle={Proceedings of the Fourth International Symposium of Chinese CHI},
  pages={1--6},
  year={2016}
}

@article{orlosky2014managing,
  title={Managing mobile text in head mounted displays: studies on visual preference and text placement},
  author={Orlosky, Jason and Kiyokawa, Kiyoshi and Takemura, Haruo},
  journal={ACM SIGMOBILE Mobile Computing and Communications Review},
  volume={18},
  number={2},
  pages={20--31},
  year={2014},
  publisher={ACM New York, NY, USA}
}

@inproceedings{ku2019peritext,
  title={PeriText: Utilizing peripheral vision for reading text on augmented reality smart glasses},
  author={Ku, Pin-Sung and Lin, Yu-Chih and Peng, Yi-Hao and Chen, Mike Y},
  booktitle={2019 IEEE Conference on Virtual Reality and 3D User Interfaces (VR)},
  pages={630--635},
  year={2019},
  organization={IEEE}
}

@inproceedings{matsuura2019readability,
  title={Readability and legibility of fonts considering shakiness of head mounted displays},
  author={Matsuura, Yuki and Terada, Tsutomu and Aoki, Tomohiro and Sonoda, Susumu and Isoyama, Naoya and Tsukamoto, Masahiko},
  booktitle={Proceedings of the 2019 ACM international symposium on wearable computers},
  pages={150--159},
  year={2019}
}

@inproceedings{gabbard2007active,
  title={Active text drawing styles for outdoor augmented reality: A user-based study and design implications},
  author={Gabbard, Joseph L and Swan, J Edward and Hix, Deborah and Kim, Si-Jung and Fitch, Greg},
  booktitle={2007 IEEE Virtual Reality Conference},
  pages={35--42},
  year={2007},
  organization={IEEE}
}

@inproceedings{road2productivity,
author = {Patel, Shiv G and Dufresne-Camaro, Charles-Olivier and Sakamoto, Yumiko and Fan, Kevin and Hasan, Khalad and Irani, Pourang},
title = {On the Road to Productivity: Investigating Text-Presentation Techniques and Audio Assistance for Non-Driving Tasks in Conditionally Automated Vehicles},
year = {2023},
isbn = {9798400709210},
publisher = {Association for Computing Machinery},
address = {New York, NY, USA},
url = {https://doi.org/10.1145/3626705.3627787},
doi = {10.1145/3626705.3627787},
booktitle = {Proceedings of the 22nd International Conference on Mobile and Ubiquitous Multimedia},
pages = {122–133},
numpages = {12},
location = {Vienna, Austria},
series = {MUM '23}
}

@inproceedings{listening-or-reading,
  title={Listening Ears or Reading Eyes: A Meta-Analysis of Reading and Listening Comprehension Comparisons},
  author={Virginia Clinton-Lisell},
  booktitle={Review of Educational Research},
  pages={543-582},
  year={2021},
  doi = {10.3102/00346543211060871}
}

@article{schiavo2021attention,
  title={Attention-driven read-aloud technology increases reading comprehension in children with reading disabilities},
  author={Schiavo, Gianluca and Mana, Nadia and Mich, Ornella and Zancanaro, Massimo and Job, Remo},
  journal={Journal of Computer Assisted Learning},
  volume={37},
  number={3},
  pages={875--886},
  year={2021},
  publisher={Wiley Online Library}
}

@article{liu2019modality,
author = {Liu, Huiyun and Cao, Shujin and Wu, Shiyu},
title = {An experimental comparison on reading comprehension effect of visual, audio and dual channels},
journal = {Proceedings of the Association for Information Science and Technology},
volume = {56},
number = {1},
pages = {716-718},
doi = {https://doi.org/10.1002/pra2.148},
url = {https://asistdl.onlinelibrary.wiley.com/doi/abs/10.1002/pra2.148},
eprint = {https://asistdl.onlinelibrary.wiley.com/doi/pdf/10.1002/pra2.148},
year = {2019}
}

@inproceedings{wobbrock2011aligned,
  title={The aligned rank transform for nonparametric factorial analyses using only anova procedures},
  author={Wobbrock, Jacob O and Findlater, Leah and Gergle, Darren and Higgins, James J},
  booktitle={Proceedings of the SIGCHI conference on human factors in computing systems},
  pages={143--146},
  year={2011}
}

@inproceedings{visual-audioHaptic,
author = {Hsieh, Yi-Ta and Orso, Valeria and Andolina, Salvatore and Canaveras, Manuela and Cabral, Diogo and Spagnolli, Anna and Gamberini, Luciano and Jacucci, Giulio},
title = {Interweaving Visual and Audio-Haptic Augmented Reality for Urban Exploration},
year = {2018},
isbn = {9781450351980},
publisher = {Association for Computing Machinery},
address = {New York, NY, USA},
url = {https://doi.org/10.1145/3196709.3196733},
doi = {10.1145/3196709.3196733},
booktitle = {Proceedings of the 2018 Designing Interactive Systems Conference},
pages = {215–226},
numpages = {12},
location = {Hong Kong, China},
series = {DIS '18}
}

@inproceedings{continuous-reading,
title = {Mobile continuous reading},
author = {Chen-Hsiang Yu},
year = {2012},
booktitle = {CHI '12 Extended Abstracts on Human Factors in Computing Systems},
doi = {10.1145/2212776.2212463},
pages={1405-1410}
}

@inproceedings{smoothing,
author = {Kumar, Manu and Klingner, Jeff and Puranik, Rohan and Winograd, Terry and Paepcke, Andreas},
title = {Improving the accuracy of gaze input for interaction},
year = {2008},
isbn = {9781595939821},
publisher = {Association for Computing Machinery},
address = {New York, NY, USA},
url = {https://doi.org/10.1145/1344471.1344488},
doi = {10.1145/1344471.1344488},
booktitle = {Proceedings of the 2008 Symposium on Eye Tracking Research \& Applications},
pages = {65–68},
numpages = {4},
location = {Savannah, Georgia},
series = {ETRA '08}
}

@inproceedings{gazePrompt,
author = {Wang, Ru and Potter, Zach and Ho, Yun and Killough, Daniel and Zeng, Linxiu and Mondal, Sanbrita and Zhao, Yuhang},
title = {GazePrompt: Enhancing Low Vision People's Reading Experience with Gaze-Aware Augmentations},
year = {2024},
isbn = {9798400703300},
publisher = {Association for Computing Machinery},
address = {New York, NY, USA},
url = {https://doi.org/10.1145/3613904.3642878},
doi = {10.1145/3613904.3642878},
booktitle = {Proceedings of the CHI Conference on Human Factors in Computing Systems},
articleno = {894},
numpages = {17},
location = {Honolulu, HI, USA},
series = {CHI '24}
}

@incollection{nasa-tlx,
title = {Development of NASA-TLX (Task Load Index): Results of Empirical and Theoretical Research},
editor = {Peter A. Hancock and Najmedin Meshkati},
series = {Advances in Psychology},
publisher = {North-Holland},
volume = {52},
pages = {139-183},
year = {1988},
booktitle = {Human Mental Workload},
issn = {0166-4115},
doi = {https://doi.org/10.1016/S0166-4115(08)62386-9},
url = {https://www.sciencedirect.com/science/article/pii/S0166411508623869},
author = {Sandra G. Hart and Lowell E. Staveland}
}

@inproceedings{reading-onSmartGlass,
author = {Rzayev, Rufat and Wo\'{z}niak, Pawe\l{} W. and Dingler, Tilman and Henze, Niels},
title = {Reading on Smart Glasses: The Effect of Text Position, Presentation Type and Walking},
year = {2018},
isbn = {9781450356206},
publisher = {Association for Computing Machinery},
address = {New York, NY, USA},
url = {https://doi.org/10.1145/3173574.3173619},
doi = {10.1145/3173574.3173619},
booktitle = {Proceedings of the 2018 CHI Conference on Human Factors in Computing Systems},
pages = {1–9},
numpages = {9},
location = {Montreal QC, Canada},
series = {CHI '18}
}

@inproceedings{klose2019text,
  title={Text presentation for augmented reality applications in dual-task situations},
  author={Klose, Elisa Maria and Mack, Nils Adrian and Hegenberg, Jens and Schmidt, Ludger},
  booktitle={2019 IEEE Conference on Virtual Reality and 3D User Interfaces (VR)},
  pages={636--644},
  year={2019},
  organization={IEEE}
}

@inproceedings{ghosh2020eyeditor,
  title={Eyeditor: Towards on-the-go heads-up text editing using voice and manual input},
  author={Ghosh, Debjyoti and Foong, Pin Sym and Zhao, Shengdong and Liu, Can and Janaka, Nuwan and Erusu, Vinitha},
  booktitle={Proceedings of the 2020 CHI Conference on Human Factors in Computing Systems},
  pages={1--13},
  year={2020}
}

@inproceedings{lucero2014notifeye,
  title={NotifEye: using interactive glasses to deal with notifications while walking in public},
  author={Lucero, Andr{\'e}s and Vetek, Akos},
  booktitle={Proceedings of the 11th conference on advances in computer entertainment technology},
  pages={1--10},
  year={2014}
}

@inproceedings{oulasvirta2005interaction,
  title={Interaction in 4-second bursts: the fragmented nature of attentional resources in mobile HCI},
  author={Oulasvirta, Antti and Tamminen, Sakari and Roto, Virpi and Kuorelahti, Jaana},
  booktitle={Proceedings of the SIGCHI conference on Human factors in computing systems},
  pages={919--928},
  year={2005}
}

@inproceedings{reading-spacing,
author = {Zhou, Chen and Fennedy, Katherine and Tan, Felicia Fang-Yi and Zhao, Shengdong and Shao, Yurui},
title = {Not All Spacings are Created Equal: The Effect of Text Spacings in On-the-go Reading Using Optical See-Through Head-Mounted Displays},
year = {2023},
isbn = {9781450394215},
publisher = {Association for Computing Machinery},
address = {New York, NY, USA},
url = {https://doi.org/10.1145/3544548.3581430},
doi = {10.1145/3544548.3581430},
booktitle = {Proceedings of the 2023 CHI Conference on Human Factors in Computing Systems},
articleno = {720},
numpages = {19},
location = {Hamburg, Germany},
series = {CHI '23}
}

@inproceedings{comparing-coordinate,
  title={Comparing world and screen coordinate systems in optical see-through head-mounted displays for text readability while walking},
  author={Fukushima, Shogo and Hamada, Takeo and Hautasaari, Ari},
  booktitle={2020 IEEE International Symposium on Mixed and Augmented Reality (ISMAR)},
  pages={649--658},
  year={2020},
  organization={IEEE}
}

@ARTICLE{hazardSnap,
  author={Zhao, Guanghan and Orlosky, Jason and Gabbard, Joseph and Kiyokawa, Kiyoshi},
  journal={IEEE Transactions on Visualization and Computer Graphics}, 
  title={HazARdSnap: Gazed-based Augmentation Delivery for Safe Information Access while Cycling}, 
  year={2023},
  volume={},
  number={},
  pages={1-10},
  doi={10.1109/TVCG.2023.3333336}}

@inproceedings{flowAR,
author = {Jo, Hye-Young and Seidel, Laurenz and Pahud, Michel and Sinclair, Mike and Bianchi, Andrea},
title = {FlowAR: How Different Augmented Reality Visualizations of Online Fitness Videos Support Flow for At-Home Yoga Exercises},
year = {2023},
isbn = {9781450394215},
publisher = {Association for Computing Machinery},
address = {New York, NY, USA},
url = {https://doi.org/10.1145/3544548.3580897},
doi = {10.1145/3544548.3580897},
booktitle = {Proceedings of the 2023 CHI Conference on Human Factors in Computing Systems},
articleno = {469},
numpages = {17},
location = {Hamburg, Germany},
series = {CHI '23}
}

@inproceedings{here&now,
author = {Zhou, Qiushi and Grebel, Louise and Irlitti, Andrew and Minaai, Julie Ann and Goncalves, Jorge and Velloso, Eduardo},
title = {Here and Now: Creating Improvisational Dance Movements with a Mixed Reality Mirror},
year = {2023},
isbn = {9781450394215},
publisher = {Association for Computing Machinery},
address = {New York, NY, USA},
url = {https://doi.org/10.1145/3544548.3580666},
doi = {10.1145/3544548.3580666},
booktitle = {Proceedings of the 2023 CHI Conference on Human Factors in Computing Systems},
articleno = {183},
numpages = {16},
location = {Hamburg, Germany},
series = {CHI '23}
}

@inproceedings{dynamicTextManagement,
author = {Orlosky, Jason and Kiyokawa, Kiyoshi and Takemura, Haruo},
title = {Dynamic text management for see-through wearable and heads-up display systems},
year = {2013},
isbn = {9781450319652},
publisher = {Association for Computing Machinery},
address = {New York, NY, USA},
url = {https://doi.org/10.1145/2449396.2449443},
doi = {10.1145/2449396.2449443},
booktitle = {Proceedings of the 2013 International Conference on Intelligent User Interfaces},
pages = {363–370},
numpages = {8},
location = {Santa Monica, California, USA},
series = {IUI '13}
}

@inproceedings{lu2020glanceable,
  title={Glanceable ar: Evaluating information access methods for head-worn augmented reality},
  author={Lu, Feiyu and Davari, Shakiba and Lisle, Lee and Li, Yuan and Bowman, Doug A},
  booktitle={2020 IEEE conference on virtual reality and 3D user interfaces (VR)},
  pages={930--939},
  year={2020},
  organization={IEEE}
}

@article{glassMessaging,
author = {Janaka, Nuwan and Gao, Jie and Zhu, Lin and Zhao, Shengdong and Lyu, Lan and Xu, Peisen and Nabokow, Maximilian and Wang, Silang and Ong, Yanch},
title = {GlassMessaging: Towards Ubiquitous Messaging Using OHMDs},
year = {2023},
issue_date = {September 2023},
publisher = {Association for Computing Machinery},
address = {New York, NY, USA},
volume = {7},
number = {3},
url = {https://doi.org/10.1145/3610931},
doi = {10.1145/3610931},
journal = {Proc. ACM Interact. Mob. Wearable Ubiquitous Technol.},
month = {sep},
articleno = {100},
numpages = {32}
}

@inproceedings{glassMail,
author = {Zhou, Chen and Yan, Zihan and Ram, Ashwin and Gu, Yue and Xiang, Yan and Liu, Can and Huang, Yun and Ooi, Wei Tsang and Zhao, Shengdong},
title = {GlassMail: Towards Personalised Wearable Assistant for On-the-Go Email Creation on Smart Glasses},
year = {2024},
isbn = {9798400705830},
publisher = {Association for Computing Machinery},
address = {New York, NY, USA},
url = {https://doi.org/10.1145/3643834.3660683},
doi = {10.1145/3643834.3660683},
booktitle = {Proceedings of the 2024 ACM Designing Interactive Systems Conference},
pages = {372–390},
numpages = {19},
location = {IT University of Copenhagen, Denmark},
series = {DIS '24}
}

@inproceedings{auralBrowsing,
author = {Yang, Tao and Ferati, Mexhid and Liu, Yikun and Rohani Ghahari, Romisa and Bolchini, Davide},
title = {Aural browsing on-the-go: listening-based back navigation in large web architectures},
year = {2012},
isbn = {9781450310154},
publisher = {Association for Computing Machinery},
address = {New York, NY, USA},
url = {https://doi.org/10.1145/2207676.2207715},
doi = {10.1145/2207676.2207715},
booktitle = {Proceedings of the SIGCHI Conference on Human Factors in Computing Systems},
pages = {277–286},
numpages = {10},
location = {Austin, Texas, USA},
series = {CHI '12}
}

@inproceedings{auralNavigation,
author = {Gross, Mikaylah and Dara, Joe and Meyer, Christopher and Bolchini, Davide},
title = {Exploring Aural Navigation by Screenless Access},
year = {2018},
isbn = {9781450356510},
publisher = {Association for Computing Machinery},
address = {New York, NY, USA},
url = {https://doi.org/10.1145/3192714.3192815},
doi = {10.1145/3192714.3192815},
booktitle = {Proceedings of the 15th International Web for All Conference},
articleno = {27},
numpages = {10},
location = {Lyon, France},
series = {W4A '18}
}

@article{rohani2016semi,
  title={Semi-aural interfaces: Investigating voice-controlled aural flows},
  author={Rohani Ghahari, Romisa and George-Palilonis, Jennifer and Gahangir, Hossain and Kaser, Lindsay N and Bolchini, Davide},
  journal={Interacting with Computers},
  volume={28},
  number={6},
  pages={826--842},
  year={2016},
  publisher={Oxford University Press}
}

@inproceedings{vrReadingUIs,
author = {Dingler, Tilman and Kunze, Kai and Outram, Benjamin},
title = {VR Reading UIs: Assessing Text Parameters for Reading in VR},
year = {2018},
isbn = {9781450356213},
publisher = {Association for Computing Machinery},
address = {New York, NY, USA},
url = {https://doi.org/10.1145/3170427.3188695},
doi = {10.1145/3170427.3188695},
booktitle = {Extended Abstracts of the 2018 CHI Conference on Human Factors in Computing Systems},
pages = {1–6},
numpages = {6},
location = {Montreal QC, Canada},
series = {CHI EA '18}
}

@ARTICLE{textReadability,
  author={Debernardis, Saverio and Fiorentino, Michele and Gattullo, Michele and Monno, Giuseppe and Uva, Antonio Emmanuele},
  journal={IEEE Transactions on Visualization and Computer Graphics}, 
  title={Text Readability in Head-Worn Displays: Color and Style Optimization in Video versus Optical See-Through Devices}, 
  year={2014},
  volume={20},
  number={1},
  pages={125-139},
  doi={10.1109/TVCG.2013.86}
}

@INPROCEEDINGS{reading-3dSurfaces,
  author={Wei, Chunxue and Yu, Difeng and Dingler, Tilman},
  booktitle={2020 IEEE Conference on Virtual Reality and 3D User Interfaces (VR)}, 
  title={Reading on 3D Surfaces in Virtual Environments}, 
  year={2020},
  volume={},
  number={},
  pages={721-728},
  doi={10.1109/VR46266.2020.00095}}

@article{sus,
author = {Brooke, John},
year = {1995},
month = {11},
pages = {},
title = {SUS: A quick and dirty usability scale},
volume = {189},
journal = {Usability Eval. Ind.}
}

@misc{Hololens,
    author  = {Microsoft},
    title   = {Microsoft HoloLens|Mixed Reality Technology for business},
    note    = {\url{https://www.microsoft.com/en-us/hololens}},
    year    = {2019}
}

@article{ginters2019augmented,
  title={Augmented reality use for cycling quality improvement},
  author={Ginters, Egils},
  journal={Procedia Computer Science},
  volume={149},
  pages={167--176},
  year={2019},
  publisher={Elsevier}
}

@inproceedings{lu2021evaluating,
  title={Evaluating the potential of glanceable ar interfaces for authentic everyday uses},
  author={Lu, Feiyu and Bowman, Doug A},
  booktitle={2021 IEEE virtual reality and 3D user interfaces (VR)},
  pages={768--777},
  year={2021},
  organization={IEEE}
}

@inproceedings{dancu2015gesture,
  title={Gesture bike: examining projection surfaces and turn signal systems for urban cycling},
  author={Dancu, Alexandru and Vechev, Velko and {\"U}nl{\"u}er, Adviye Ay{\c{c}}a and Nilson, Simon and Nygren, Oscar and Eliasson, Simon and Barjonet, Jean-Elie and Marshall, Joe and Fjeld, Morten},
  booktitle={Proceedings of the 2015 international conference on interactive tabletops \& surfaces},
  pages={151--159},
  year={2015}
}

@inproceedings{chatterjee2020smarthelm,
  title={Smarthelm: Towards multimodal detection of attention in an outdoor augmented reality biking scenario},
  author={Chatterjee, Sromona and Scheck, Kevin and K{\"u}ster, Dennis and Putze, Felix and Moturu, Harish and Schering, Johannes and G{\'o}mez, Jorge Marx and Schultz, Tanja},
  booktitle={Companion Publication of the 2020 International Conference on Multimodal Interaction},
  pages={426--432},
  year={2020}
}

@inproceedings{matviienko2022bikear,
  title={BikeAR: Understanding cyclists’ crossing decision-making at uncontrolled intersections using Augmented Reality},
  author={Matviienko, Andrii and M{\"u}ller, Florian and Sch{\"o}n, Dominik and Seesemann, Paul and G{\"u}nther, Sebastian and M{\"u}hlh{\"a}user, Max},
  booktitle={Proceedings of the 2022 CHI Conference on Human Factors in Computing Systems},
  pages={1--15},
  year={2022}
}

@inproceedings{Li2025situationadapt,
author = {Li, Zhipeng and Gebhardt, Christoph and Inglin, Yves and Steck, Nicolas and Streli, Paul and Holz, Christian},
title = {SituationAdapt: Contextual UI Optimization in Mixed Reality with Situation Awareness via LLM Reasoning},
year = {2024},
isbn = {9798400706288},
publisher = {Association for Computing Machinery},
address = {New York, NY, USA},
url = {https://doi.org/10.1145/3654777.3676470},
doi = {10.1145/3654777.3676470},
booktitle = {Proceedings of the 37th Annual ACM Symposium on User Interface Software and Technology},
articleno = {43},
numpages = {13},
location = {Pittsburgh, PA, USA},
series = {UIST '24}
}

@inproceedings{Lindlbauer2019context,
author = {Lindlbauer, David and Feit, Anna Maria and Hilliges, Otmar},
title = {Context-Aware Online Adaptation of Mixed Reality Interfaces},
year = {2019},
isbn = {9781450368162},
publisher = {Association for Computing Machinery},
address = {New York, NY, USA},
url = {https://doi.org/10.1145/3332165.3347945},
doi = {10.1145/3332165.3347945},
booktitle = {Proceedings of the 32nd Annual ACM Symposium on User Interface Software and Technology},
pages = {147–160},
numpages = {14},
location = {New Orleans, LA, USA},
series = {UIST '19}
}

@article{Huang2024Reading,
    author = {Yi-Jheng Huang and Jing-Cheng Lin and Suiang-Shyan Lee and Bo-Jheng Wu},
    title = {Reading and Walking with Smart Glasses: Effects of Display and Control Modes on Safety},
    journal = {International Journal of Human–Computer Interaction},
    volume = {40},
    number = {23},
    pages = {7875--7891},
    year = {2024},
    publisher = {Taylor \& Francis},
    doi = {10.1080/10447318.2023.2276529},
    URL = {https://doi.org/10.1080/10447318.2023.2276529},
    eprint = {https://doi.org/10.1080/10447318.2023.2276529}
}

@inproceedings{Rzayev2020Effects,
    author = {Rzayev, Rufat and Korbely, Susanne and Maul, Milena and Schark, Alina and Schwind, Valentin and Henze, Niels},
    title = {Effects of Position and Alignment of Notifications on AR Glasses during Social Interaction},
    year = {2020},
    isbn = {9781450375795},
    publisher = {Association for Computing Machinery},
    address = {New York, NY, USA},
    url = {https://doi.org/10.1145/3419249.3420095},
    doi = {10.1145/3419249.3420095},
    booktitle = {Proceedings of the 11th Nordic Conference on Human-Computer Interaction: Shaping Experiences, Shaping Society},
    articleno = {30},
    numpages = {11},
    location = {Tallinn, Estonia},
    series = {NordiCHI '20}
}

@inproceedings{Koelle2015Dont,
author = {Koelle, Marion and Kranz, Matthias and M\"{o}ller, Andreas},
title = {Don't look at me that way! Understanding User Attitudes Towards Data Glasses Usage},
year = {2015},
isbn = {9781450336529},
publisher = {Association for Computing Machinery},
address = {New York, NY, USA},
url = {https://doi.org/10.1145/2785830.2785842},
doi = {10.1145/2785830.2785842},
booktitle = {Proceedings of the 17th International Conference on Human-Computer Interaction with Mobile Devices and Services},
pages = {362–372},
numpages = {11},
location = {Copenhagen, Denmark},
series = {MobileHCI '15}
}

@inproceedings{Lu2023InTheWild,
author = {Lu, Feiyu and Pavanatto, Leonardo and Bowman, Doug A.},
title = {In-the-Wild Experiences with an Interactive Glanceable AR System for Everyday Use},
year = {2023},
isbn = {9798400702815},
publisher = {Association for Computing Machinery},
address = {New York, NY, USA},
url = {https://doi.org/10.1145/3607822.3614515},
doi = {10.1145/3607822.3614515},
booktitle = {Proceedings of the 2023 ACM Symposium on Spatial User Interaction},
articleno = {11},
numpages = {9},
location = {Sydney, NSW, Australia},
series = {SUI '23}
}

@incollection{hart1988development,
  title={Development of NASA-TLX (Task Load Index): Results of empirical and theoretical research},
  author={Hart, Sandra G and Staveland, Lowell E},
  booktitle={Advances in psychology},
  volume={52},
  pages={139--183},
  year={1988},
  publisher={Elsevier}
}

@article{bangor2008empirical,
  title={An empirical evaluation of the system usability scale},
  author={Bangor, Aaron and Kortum, Philip T and Miller, James T},
  journal={Intl. Journal of Human--Computer Interaction},
  volume={24},
  number={6},
  pages={574--594},
  year={2008},
  publisher={Taylor \& Francis}
}

@inproceedings{zhao2024rear,
  title={ReAR Indicators: Peripheral Cycling Indicators for Rear-Approaching Hazards},
  author={Zhao, Guanghan and Hu, Xiaodan and Orlosky, Jason and Kiyokawa, Kiyoshi},
  booktitle={Proceedings of the 2024 International Conference on Advanced Visual Interfaces},
  pages={1--9},
  year={2024}
}
\end{document}